\documentclass[fleqn,usenatbib]{mnras}
\usepackage{newtxtext,newtxmath}
\usepackage[T1]{fontenc}
\usepackage{ae,aecompl}
\usepackage{graphicx}
\usepackage{float}
\usepackage{caption}
\usepackage{multicol}
\usepackage{multirow}
\usepackage{longtable}

\title[Radio spectra of NLS1s with ATCA and VLASS]{Radio spectra of narrow-line Seyfert 1 galaxies observed with ATCA and VLASS}

\author[Sina Chen et al.]{
Sina Chen, $^{1}$ \thanks{E-mail: sina.chen@campus.technion.ac.il}
Jamie B. Stevens, $^{2}$
Philip G. Edwards, $^{2}$
Ari Laor, $^{1}$
Minfeng Gu, $^{3}$
\newauthor
Marco Berton, $^{4}$
Emilia J{\"a}rvel{\"a}, $^{5}$
Preeti Kharb, $^{6}$
Ehud Behar, $^{1}$
and Renzhi Su. $^{3}$
\\
$^{1}$ Physics Department, Technion, Haifa 32000, Israel \\
$^{2}$ CSIRO Astronomy and Space Science, Australian Telescope National Facility, Epping, New South Wales 1710, Australia \\
$^{3}$ Key Laboratory for Research in Galaxies and Cosmology, Shanghai Astronomical Observatory, Chinese Academy of Sciences, 80 Nandan Road, Shanghai \\ 200030, China \\
$^{4}$ European Southern Observatory (ESO), Al{\'o}nso de Cordova 3107, Casilla 19, Santiago 19001, Chile \\
$^{5}$ European Space Agency, European Space Astronomy Centre, C/Bajo el Castillo s/n, 28692 Villanueva de la Ca{\~n}ada, Madrid, Spain \\
$^{6}$ National Centre for Radio Astrophysics - Tata Institute of Fundamental Research, Post Bag 3, Ganeshkhind, Pune 411007, India \\
}

\date{Accepted XXX, Received YYY, in original form ZZZ.}

\pubyear{2022}

\begin{document}
\label{firstpage}
\pagerange{\pageref{firstpage}--\pageref{lastpage}}
\maketitle

\begin{abstract}

We present radio spectral analyses for a sample of 29 radio-quiet (RQ) and three radio-loud (RL) narrow-line Seyfert 1 galaxies (NLS1s) detected with the Australia Telescope Compact Array at both 5.5~GHz and 9.0~GHz.
The sample is characterized by $L_{\rm bol}/L_{\rm Edd} > 0.15$.
The radio slopes in 25 of the 29 RQ NLS1s are steep ($\alpha_{5.5-9.0} < -0.5$), as found in earlier studies of RQ high $L_{\rm bol}/L_{\rm Edd}$ AGN.
This steep radio emission may be related to AGN-driven outflows, which are likely more prevalent in high $L_{\rm bol}/L_{\rm Edd}$ AGN.
In two of the three RL NLS1s, the radio slopes are flat or inverted ($\alpha_{5.5-9.0} > -0.5$), indicating a compact optically-thick source, likely a relativistic jet.
Archival data at 3.0~GHz, 1.4~GHz, and 0.843~GHz are also compiled, yielding a sample of 17 NLS1s detected in three bands or more.
In nine objects, the radio spectra flatten at lower frequencies, with median slopes of $\alpha_{5.5-9.0} = -1.21 \pm 0.17$, flattening to $\alpha_{3.0-5.5} = -0.97 \pm 0.27$, and to $\alpha_{1.4-3.0} = -0.63 \pm 0.16$.
A parabolic fit suggests a median spectral turnover of $\sim$~1~GHz, which implies synchrotron self-absorption in a source with a size of only a fraction of 1~pc, possibly a compact wind or a weak jet.
Two objects show significant spectral steepening to $\alpha < -2$ above 3~GHz or 5~GHz, which may suggest relic emission from past ejection of radio emitting plasma, of the order of a few years to a few decades ago.
Finally, two objects present a single spectral slope consistent with star-forming activity.

\end{abstract}


\section{Introduction}

Narrow-line Seyfert 1 galaxies (NLS1s) are a well-known subclass of active galactic nuclei (AGN). They are defined by optical spectral properties with a full width at half maximum of broad H$\beta$ emission lines of FWHM(H$\beta$) $<$ 2000 km s$^{-1}$ and a flux ratio of [O III]$\lambda$5007 / H$\beta$ $<$ 3 \citep{Osterbrock1985,Goodrich1989}. Compared to normal broad-line Seyfert 1 galaxies, NLS1s exhibit strong Fe~II multiplets emission \citep{Boroson1992}, frequently observed blueshifted line profiles \citep{Leighly2004,Boroson2005}, steep soft X-ray spectra \citep{Laor1994,Boller1996,Wang1996,Leighly1999a}, and rapid X-ray variability \citep{Pounds1995,Leighly1999b}.

The narrowness of Balmer lines in NLS1s is commonly interpreted as a low rotational velocity of the gas in the broad-line region (BLR) around a central black hole (BH) typically in the range of $M_{\rm BH} \sim 10^6$--$10^7 M_{\odot}$, which is relatively low-mass compared to normal Seyfert 1 galaxies \citep{Peterson2011,Jarvela2015,Cracco2016,Wang2016,Chen2018a}.
They accrete at high rates, from 0.1 Eddington ratio to close or even higher than their Eddington limit \citep{Boroson1992}.
Some studies found that NLS1s lie below the BH-bulge mass relation and the BH mass vs.\ stellar velocity dispersion relation \citep[e.g.][]{Mathur2001,Zhou2006}, but some disagree \citep[e.g.][]{Woo2015}.
Another interpretation is that the narrowness of Balmer lines is just an inclination effect caused by a pole-on orientation of a disk-like BLR \citep{Decarli2008}.
However, studies of their host galaxy environments \citep{Boroson1992,Kollatschny2011,Jarvela2018,Olguiniglesias2020}, and reverberation mapping campaigns of individual NLS1s \citep{Du2014,Wang2016,Du2019}, reveal that the differences between the normal Seyfert 1 and the NLS1 populations should be intrinsic rather than just an orientation effect.

The majority of NLS1s are radio-quiet (RQ) with a radio loudness of $R_{\rm L} < 10$, and only a fraction ($\sim 7\%$) of NLS1s are radio-loud (RL) with $R_{\rm L} > 10$ \citep{Komossa2006}, where $R_{\rm L}$ is defined as a ratio of the 5 GHz radio flux density of the optical B-band flux density $R_{\rm L} = S_{5\rm{GHz}} / S_{4400\text{\AA}}$ \citep{Kellermann1989}.
Multi-wavelength studies show that RL NLS1s are likely to harbor relativistic jets, which contribute emission in radio, optical, partially X-ray, and in some cases $\gamma$-ray. They have similar behavior, e.g.\ the spectral energy distribution (SED), to flat spectrum radio quasars \citep[FSRQs,][]{Yuan2008,Foschini2015,Foschini2017,Foschini2020}. Radio observations of RL NLS1s with the Very Long Baseline Array (VLBA) show that they exhibit pc-scale jets, compact morphologies, flat or inverted spectra, high brightness temperatures, and a significant degree of polarization \citep{Abdo2009a,Abdo2009b,Gu2015,Lister2018}.
Unlike FSRQs which show these jet features on kpc scales, NLS1s rarely harbor extended radio emission and only exhibit an unresolved core on kpc scales \citep{Berton2018,Chen2020,Jarvela2021b}.
Radio monitoring campaigns also reveal the presence of fast and intense variability which is a common feature of relativistic jets \citep{Foschini2012a,Angelakis2015,Fuhrmann2016,Lahteenmaki2017,Lahteenmaki2018}. In some cases, apparent superluminal motion is observed as well \citep{Lister2013,Lister2016,Lister2019}. In addition, the discovery of $\gamma$-ray emission from RL NLS1s detected by the \textit{Fermi} Large Area Telescope (LAT) confirms the presence of relativistic beamed jets in this type of AGN along with blazars and radio galaxies \citep{Abdo2009a,Abdo2009b,Abdo2009c,Foschini2011b,Foschini2015,Yao2015,Liao2015,D'Ammando2015,D'Ammando2016,Paliya2016,Berton2017,Paliya2018,Lahteenmaki2018,Yao2019,Rakshit2021}.
However, star formation (SF) can also contribute to the observed radio emission in RL NLS1s \citep{Caccianiga2015}.

In RQ NLS1s, various radio emission mechanisms may be involved, including SF, AGN-driven wide-angle winds interacting with ambient medium, collimated low-power jets, and accretion disk coronal emission \citep{Panessa2019}. Radio observations can be used to probe these ongoing physical processes.
SF can produce radio emission via synchrotron radiation from supernova remnants and cosmic rays, and via free-free emission from H~II regions \citep{Condon1992}.
The SF radio morphology is quite diffuse (on host galaxy scale) and circular around the nucleus with a low brightness temperature typically $T_{\rm B} < 10^5$~K \citep{Christopoulou1997,Orienti2015}.
The SF radio spectrum remains optically-thin with $S_{\nu} \propto \nu^{-0.8 \pm 0.4}$ from $\sim$ 10 GHz down to $\sim$ a few hundreds MHz \citep{Magnelli2015,CalistroRivera2017,Gim2019,An2021}.
Jets and winds are expected to produce linear structures from pc to kpc scales \citep{Berton2018,Chen2020,Yao2021}.
The spectral slope is flat if it is dominated by the central core or steep if the extended emission dominates.
It is currently hard to distinguish whether the radio emission originates from jets or winds, based on the radio spectra and morphology.
High-resolution observations on pc scales, e.g.\ VLBA and Very Long Baseline Interferometry (VLBI), are needed to clarify whether the radio emission is collimated or not and if there is a jet base with high $T_{\rm B}$.
If the radio emission is of coronal origin, it is expected to be compact and optically-thick on sub-pc scales.
The base of the jet/wind may physically coincide with the corona where the particles are produced by non-thermal processes with $T_{\rm B} > 10^7$~K \citep{Blundell1998,Merloni2002,King2017}.
However, at present, it is often hard to separate the radio emission from these different sources, in particular when the radio features are unresolved or marginally resolved.

In order to study the radio properties of NLS1s, we obtained new observations with the Australia Telescope Compact Array (ATCA) and made use of the Very Large Array Sky Survey (VLASS) images. This paper presents the main results of this survey and is organized as follows. In Section 2 we describe the sample selection and observations, in Section 3 we present the data analysis, in Section 4 we discuss the main results, and in Section 5 we provide the summary. Throughout this work, we adopt a standard $\Lambda$CDM cosmology with a Hubble constant $H_{0}$ = 70~km~s$^{-1}$~Mpc$^{-1}$, $\Omega_{\Lambda}$ = 0.73 and $\Omega_{\rm{M}}$ = 0.27 \citep{Komatsu2011}. We assume the flux density and spectral index convention of $S_{\nu} \propto \nu^{\alpha}$.

\section{Sample selection and observations}

The sample is selected from \citet{Chen2018a} who presented a catalog of NLS1s in the southern hemisphere. There are 168 \footnote{There are 167 NLS1s in \citet{Chen2018a}. We included one more source IRAS 13224$-$3809 as it was classified as NLS1 in \citet{Boller1993} and also detected in the 6dFGS.} NLS1s (12 RL and 156 RQ) classified according to their optical spectra from the Six-degree Field Galaxy Survey \citep[6dFGS,][]{Jones2009}.
Sixty-two sources were observed with the Karl G. Jansky Very Large Array (VLA) C-configuration between November 2018 and February 2019, and 49 of these 62 objects were detected at 5.5~GHz with flux densities ranging from 0.05~mJy to 21.28~mJy \citep{Chen2020}.
Additional 21 objects were included in the above VLA proposal, but were not observed. The declination ($\delta$) is north of $-25 \degr$, which is not suitable to observe with ATCA.
Excluding these, we observed the remaining 85 NLS1s using the ATCA in the 4cm band.
The redshifts of these 85 objects range from 0.02 to 0.50.
Thirty-nine sources at $\delta$ between $-25 \degr$ and $-40 \degr$ were observed with the 6km array between April and October 2020 (project code: C3329), and 46 sources at $\delta$ south of $-40 \degr$ were observed with the 750m array in September 2018 (project code: CX414).

Observations were carried out at two 2GHz-wide intermediate frequencies of 4.5--6.5~GHz centered at 5.5~GHz and of 8.0--10.0~GHz centered at 9.0~GHz.
The resolutions are $\sim$ 3~arcsec for the 6km array and $\sim$ 20~arcsec for the 750m array.
PKS~1934$-$63 was used as the primary flux density calibrator and was observed for 10 minutes at each frequency and at each epoch.
The exposure time is about 20--80 minutes for each target yielding image sensitivities of 0.02--0.08~mJy~beam$^{-1}$. The total integrated time is about 88 hours. Details are reported in Table \ref{observation}.

The data reduction was carried out with \textit{Miriad} which is a radio interferometry data reduction package of particular interest to users of the ATCA \citep{Sault1995}. The data was flagged if they were affected by radio frequency interference (RFI). After that, calibrations of bandpass, gain, and flux density were carried out. Images for an individual source centered at frequencies of 5.5~GHz and 9.0~GHz were created. The standard data reduction process is described in the ATCA Users Guide page \footnote{\url{https://www.narrabri.atnf.csiro.au/observing/users_guide/html/atug.html}}. We modeled the source with a Gaussian fit on the image plane and deconvolved it from the beam, to recover the source size and its position angle, and the flux densities at both frequencies.

Forty-two sources are detected in the ATCA observations including 24 objects with the 6km array and 18 objects with the 750m array.
Among them, nine objects are only detected at 5.5~GHz and one object is only detected at 9.0~GHz.
We assume a 3$\sigma$ level as an upper limit of their flux densities.
Forty objects are unresolved in the current ATCA observations, and do not have extended emission on arcsec scales.
Some objects have very elongated beam because they were only observed for one scan or cut due to RFI.
Their maps are not shown in here because their morphology usually is just an unresolved or a very elongated core.
In case the source size is too small to be determined, we adopt half of beam size as an upper limit of the source size. The source sizes are reported in Table \ref{size}, typically from sub-kpc to a few kpc.
The flux densities range from 0.13~mJy to 167.7~mJy at 5.5~GHz and from 0.11~mJy to 151.7~mJy at 9.0~GHz.
The estimated uncertainty for individual ATCA flux density measurement is about 5\% \citep{Murphy2010}.
We study these 42 objects hereafter.

\begin{figure*}
\centering
\includegraphics[width=\textwidth, trim={0cm 1cm 0cm 1cm}, clip]{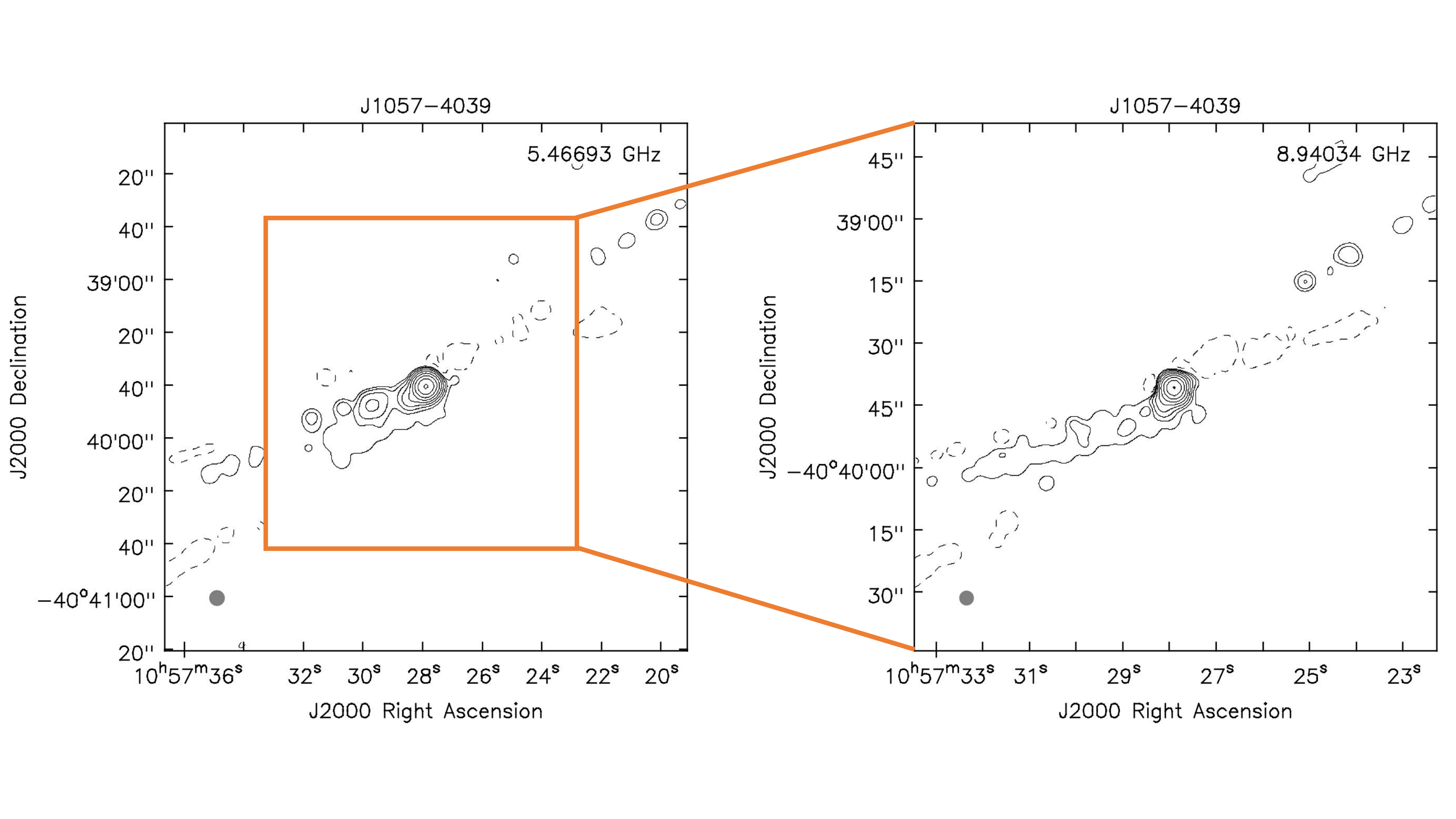}
\includegraphics[width=\textwidth, trim={0cm 1cm 0cm 1cm}, clip]{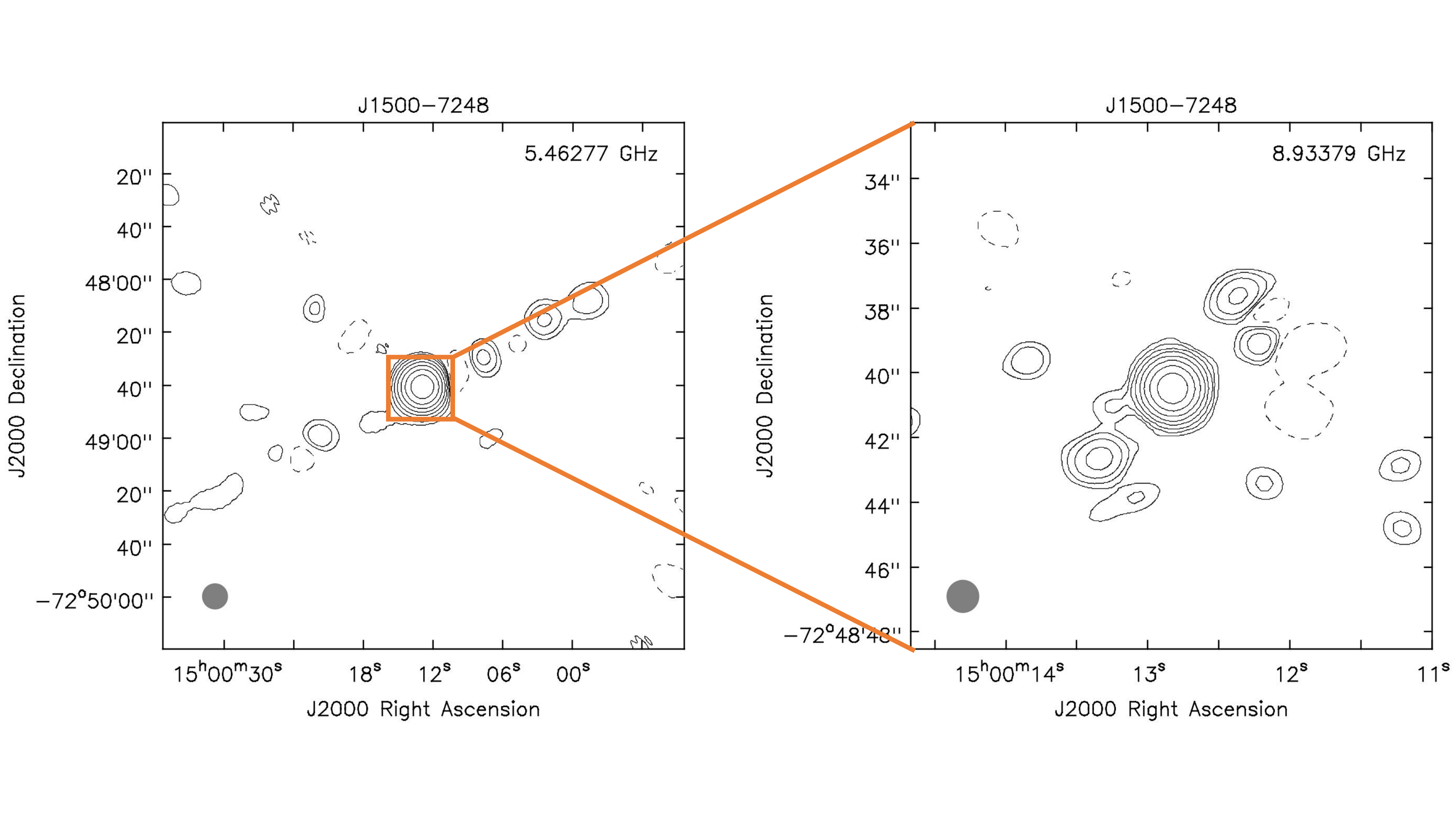}
\caption{The radio maps of the two RL NLS1s exhibiting strong extended structures.
Upper-left and upper-right panels are J1057$-$4039 at 5.5~GHz and 9.0~GHz respectively.
RMS = 0.2~mJy~beam$^{-1}$, contour levels at $-$3, 3 $\times$ 2$^n$, $n \in$ [0,8]. The beam sizes are smoothed to 5.7~arcsec $\times$ 5.7~arcsec at 5.5~GHz and to 3.4~arcsec $\times$ 3.4~arcsec at 9.0~GHz.
Lower-left and lower-right panels are J1500$-$7248 at 5.5~GHz and 9.0~GHz respectively.
RMS = 0.05~mJy~beam$^{-1}$, contour levels at $-$3, 3 $\times$ 2$^n$, $n \in$ [0,7]. The beam sizes are smoothed to 9.6~arcsec $\times$ 9.6~arcsec at 5.5~GHz and to 1.0~arcsec $\times$ 1.0~arcsec at 9.0~GHz.}
\label{maps}
\end{figure*}

Two RL objects (see Section 3), J1057$-$4039 and J1500$-$7248, exhibit strong extended structures at both 5.5~GHz and 9.0~GHz.
Their radio maps are shown in Fig.\ref{maps}.
We smoothed the original beam, which is very elongated, to a circle one in order to clarify the structure.
The morphology of J1057$-$4039 clearly shows a central core, plus an extended linear structure to the south-east of the core at both frequencies, which is likely associated with a relativistic jet (see Section 4.1.1).
In J1500$-$7248, since the smoothed beam size of the 5.5~GHz image is larger than that of the 9.0~GHz image, the 9.0~GHz image generally shows the structure of the unresolved central core in the 5.5~GHz image.
We see that this object also exhibits extended structures at both sides, which is possibly related to a two-sided jet (see Section 4.1.1).

We checked the VLASS image archive for the 42 ATCA-detected NLS1s.
VLASS is a radio continuum survey which is carried out by the National Radio Astronomy Observatory (NRAO) in S-band at 2--4~GHz and is covering $\delta \geq -40 \degr$.
The angular resolution is $\sim$ 2.5~arcsec which is comparable to the ATCA observations. The VLASS Epoch 1 observations have been completed recently (2017--2019), and the preliminary quick-look images have been made publicly available on the NRAO archive \footnote{\url{https://archive-new.nrao.edu/vlass/quicklook/}}. Each epoch of VLASS images reaches a 1$\sigma$ sensitivity of $\sim$ 0.12~mJy~beam$^{-1}$ and a flux uncertainty up to $\sim$ 20\% \citep{Lacy2020}. Despite the quick-look images not being the final data products, they are useful for identifying sources and measuring flux densities.
We searched objects according to their coordinates, and downloaded the VLASS images from the data archive.
The measurements were carried out with the Common Astronomy Software Applications (CASA).
There are 15 sources detected with VLASS, and the flux densities range from 0.39~mJy to 5.02~mJy at 3.0~GHz.
Ten sources are in the VLASS sky coverage but are undetected by VLASS.
We use a 3$\sigma$ flux level of 0.36~mJy as their upper limits.
The other 17 sources are outside the VLASS sky coverage.

We further cross-matched the catalogs of the Faint Images of the Radio Sky at Twenty-Centimeters \citep[FIRST,][]{Helfand2015} and the NRAO VLA Sky Survey \citep[NVSS,][]{Condon1998} within a search radius of 30~arcsec for the 42 sources.
No source is detected by FIRST mainly due to the sky coverage of the survey.
Eight sources are detected by NVSS with flux densities about 2.2--9.6~mJy at 1.4~GHz.
They are all detected by VLASS and ATCA.
NVSS has a resolution of 45~arcsec which is much lower than that of VLASS and ATCA.
The 1$\sigma$ sensitivity is $\sim$ 0.45~mJy~beam$^{-1}$, we thus use a 3$\sigma$ flux level of 1.35~mJy as an upper limit for those 17 objects within the NVSS sky coverage but non-detected with NVSS.
The other 17 sources are outside the NVSS sky coverage.

We additionally searched the catalog of the Sydney University Molonglo Sky Survey \citep[SUMSS,][]{Mauch2003,Mauch2008} and the second epoch Molonglo Galactic Plane Survey \citep[MGPS-2,][]{Murphy2007} within a radius of 30~arcsec for the 42 sources.
These two surveys together cover the whole southern sky with $\delta \leq -30 \degr$.
The resolution is 45~arcsec which is comparable to NVSS.
The sensitivities are $\sim$ 6~mJy~beam$^{-1}$ at $\delta \leq -50 \degr$ and $\sim$ 10~mJy~beam$^{-1}$ at $\delta > -50 \degr$.
Seven sources are detected by SUMSS with flux densities of 6.8--437.5~mJy at 843~MHz.
Two objects are outside their sky coverage.
The other 33 objects have upper limits of 6~mJy if $\delta \leq -50 \degr$ and of 10~mJy if $\delta > -50 \degr$.

In total, 42 objects are studied below. Two sources are detected in five bands, six sources are detected in four bands, nine sources are detected in three bands, 18 sources are detected in two bands, and seven sources are detected in one band.
The flux densities detected by these surveys are listed in Table \ref{flux}.

\section{Data analysis}

We calculated the radio loudness according to the flux ratio between the radio at 5~GHz and the optical at 4400~$\text{\AA}$, where the flux density at 4400~$\text{\AA}$ is derived from the B-band magnitude in the 6dFGS catalog \citep{Jones2009}.
Three sources are RL, and the other 39 sources are RQ.
The radio loudness may depend on whether the radio and optical observations include the whole galaxy (both AGN and host) or only the nucleus \citep{Ho2001b,Kharb2014} and whether the radio emission is beamed or not. Despite the boundary of $R_{\rm L}$ = 10 being arbitrary, it can give us an idea of the fraction of radio flux density with respect to optical one and make our results comparable to other studies using this parameter.

In addition, we measured the spectral index between 5.5~GHz and 9.0~GHz, $\alpha_{5.5-9.0}$, by fitting the spectrum with a power-law using the definition
\begin{equation}
\alpha = \frac{\log(S_2/S_1)}{\log(\nu_2/\nu_1)}
\label{eqa}
\end{equation}
where $S_1$ and $S_2$ are the flux densities at the observing frequencies $\nu_1$ = 5.5~GHz and $\nu_2$ = 9.0~GHz respectively. Six sources have a flat radio spectrum ($\alpha_{5.5-9.0} > -0.5$), but in only three of them at a level of more than 1$\sigma$, and 26 sources have a steep radio spectrum ($\alpha_{5.5-9.0} < -0.5$).
The other ten objects are only detected at either 5~GHz or 9~GHz and thus only have an upper or lower limit on the slope.
The uncertainties were evaluated by a Python uncertainties package \footnote{\url{https://pythonhosted.org/uncertainties/}} following the error propagation analysis.
The same method was also used to calculate the spectral indices at 3.0--5.5~GHz ($\alpha_{3.0-5.5}$), 1.4--3.0~GHz ($\alpha_{1.4-3.0}$), 0.8--1.4~GHz ($\alpha_{0.8-1.4}$), and 0.8--5.5~GHz ($\alpha_{0.8-5.5}$), when they are available.
The median spectral indices are $\alpha_{5.5-9.0} = -0.99 \pm 0.20$, $\alpha_{3.0-5.5} = -0.95 \pm 0.29$, $\alpha_{1.4-3.0} = -0.72 \pm 0.23$, $\alpha_{0.8-1.4} = -0.53 \pm 0.27$, and $\alpha_{0.8-5.5} = -0.81 \pm 0.04$ when they are detected.
We note that the spectral indices have large uncertainties due to the errors in the flux densities.
All the spectral slopes are reported in Table \ref{slope}.

The FWHM, flux, and luminosity of H$\beta$ line are from \citet{Chen2018a}.
The previous BH mass in \citet{Chen2018a} was measured via the 5100~$\text{\AA}$ continuum luminosity and the H$\beta$ line dispersion, which is sensitive to the continuum placement. This method introduces large errors especially when the signal-to-noise (S/N) ratio of the optical spectra is low.
We thus re-estimated the virial BH mass based on only the H$\beta$ line \citep{Greene2005b}
\begin{equation}
M_{\rm BH} = (3.6 \pm 0.2) \times 10^6 \left( \frac{L_{\rm{H}\beta}}{10^{42} \, \rm{erg} \, \rm{s}^{-1}} \right)^{0.56 \pm 0.02} \left( \frac{\rm{FWHM}_{\rm{H}\beta}}{10^3 \, \rm{km} \, \rm{s}^{-1}} \right)^2 M_{\odot}
\end{equation}
This method is based on the empirical relations that the BLR size and the Balmer line luminosity both scale almost linearly with the 5100~$\text{\AA}$ continuum luminosity.

\begin{figure}
\centering
\includegraphics[width=.5\textwidth]{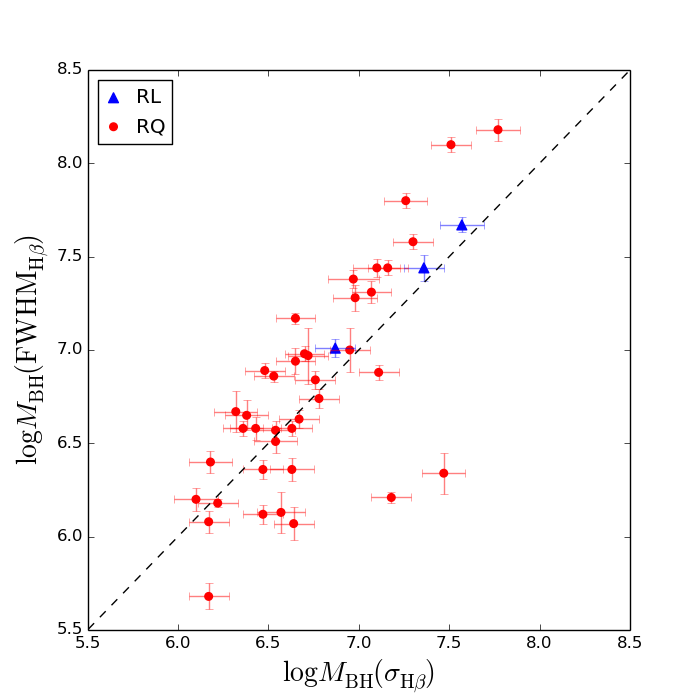}
\caption{The BH mass calculated based on the FWHM and luminosity of the H$\beta$ line, against the BH mass from \citet{Chen2018a} based on the continuum luminosity and the H$\beta$ line dispersion. The blue triangles and red circles represent RL and RQ objects. The black dashed line is the 1:1 ratio line.}
\label{compare}
\end{figure}

We plot the re-calculated BH mass against the BH mass from \citet{Chen2018a} in Fig.\ref{compare}. The majority of sources have slightly larger BH mass using the FWHM of H$\beta$ line than that using the H$\beta$ line dispersion.
There are a few outliers with a discrepancy between the BH mass measured via $\sigma_{\rm{H}\beta}$ and that calculated via FWHM$_{\rm{H}\beta}$ up to about an order of magnitude.
This is because of the different values of the H$\beta$ FWHM and the H$\beta$ line dispersion derived from the same optical spectrum, and the different relations we used to compute the BH mass.
The re-calculated BH masses are consistent with the measurements by \citet{Durre2021} in ten overlapping objects, but with a systematic offset due to the different BH mass relation we used.

Following the error propagation in these calculations, the errors of the BH mass are about 0.1--0.2~dex.
A reverberation mapping study of supermassive BH with high accretion rates in AGN by \citet{Du2018} found that the H$\beta$ time lags in AGN with high Eddington ratios are shorter than the values predicted by the canonical $R_{\rm{H}\beta}$-$L_{5100\text{\AA}}$ relation by factors of $\sim$~2--6, depending on the accretion rate.
This indicates that the BLR size and thus the BH mass, may be overestimated by $\sim$~0.3--0.8~dex if we used the canonical $R_{\rm{H}\beta}$-$L_{5100\text{\AA}}$ relation.
Thus the uncertainties of the BH mass may be as large as $\sim$~0.5--1.0~dex.

The 5100~$\text{\AA}$ continuum luminosity was measured from the optical spectrum, which was first corrected for galactic extinction, then shifted to the rest frame (source frame), and finally subtracted the host galaxy.
These values are taken from \citet{Chen2018a}.
We made a correction to the 5100~$\text{\AA}$ luminosity by multiplying (1+$z$), which was omitted in \citet{Chen2018a}.
The Eddington ratio $L_{\rm bol}/L_{\rm Edd}$ was re-calculated using these new values, where $L_{\rm bol} = 9 \times L_{5100\text{\AA}}$ and $L_{\rm Edd} = 1.3 \times 10^{38} (M_{\rm BH}/M_{\odot})$.
The Fe~II multiplets were fitted to the continuum-subtracted spectrum using the procedure described in \citet{Kovacevic2010,Shapovalova2012} \footnote{\url{http://servo.aob.rs/FeII_AGN/}}.
The flux of Fe~II ($\lambda$4570) was measured by integrating the Fe~II spectral interval between 4434~$\text{\AA}$ and 4684~$\text{\AA}$.
The flux error was evaluated by the Monte Carlo method. This approach involves varying the line profile with a random noise proportional to the RMS measured in the continuum (5050--5150~$\text{\AA}$), then integrating the line profile as previously mentioned, and repeating the same process 100 times, to estimate the standard deviation. In this way, we obtained a 1$\sigma$ error of the Fe~II ($\lambda$4570) flux. We then calculated the flux ratio of Fe~II / H$\beta$ $R_{\rm{Fe~II}/\rm{H}\beta}$.
Table \ref{others} presents the radio loudness, B-band magnitude, FWHM of H$\beta$ line, flux ratio of Fe~II / H$\beta$, optical continuum luminosity, BH mass, and Eddington ratio.

\section{Discussions}

\subsection{The radio spectra}

\begin{figure*}
\centering
\includegraphics[width=.43\textwidth]{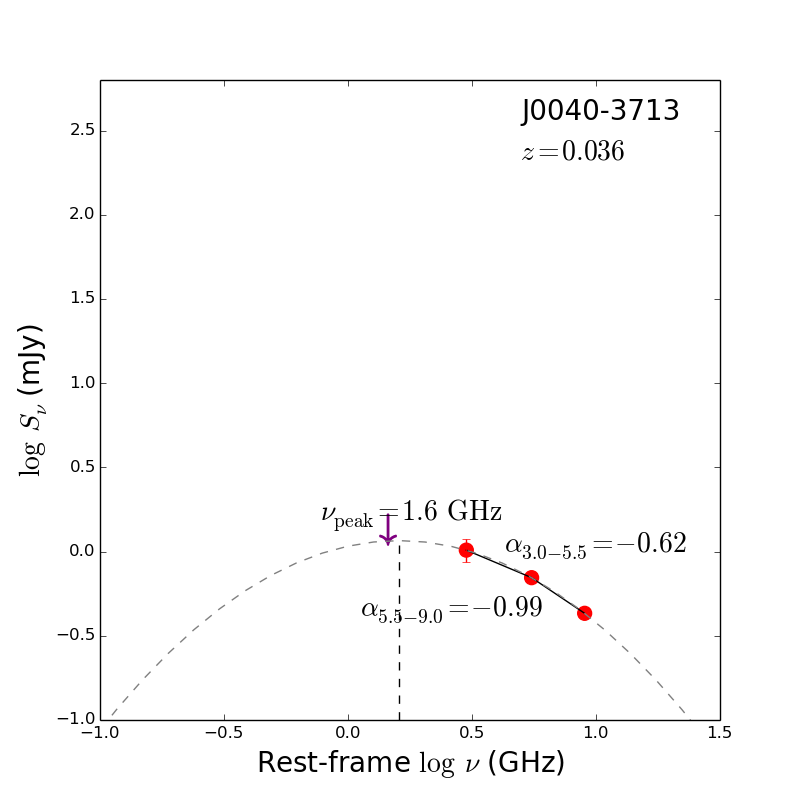}
\includegraphics[width=.43\textwidth]{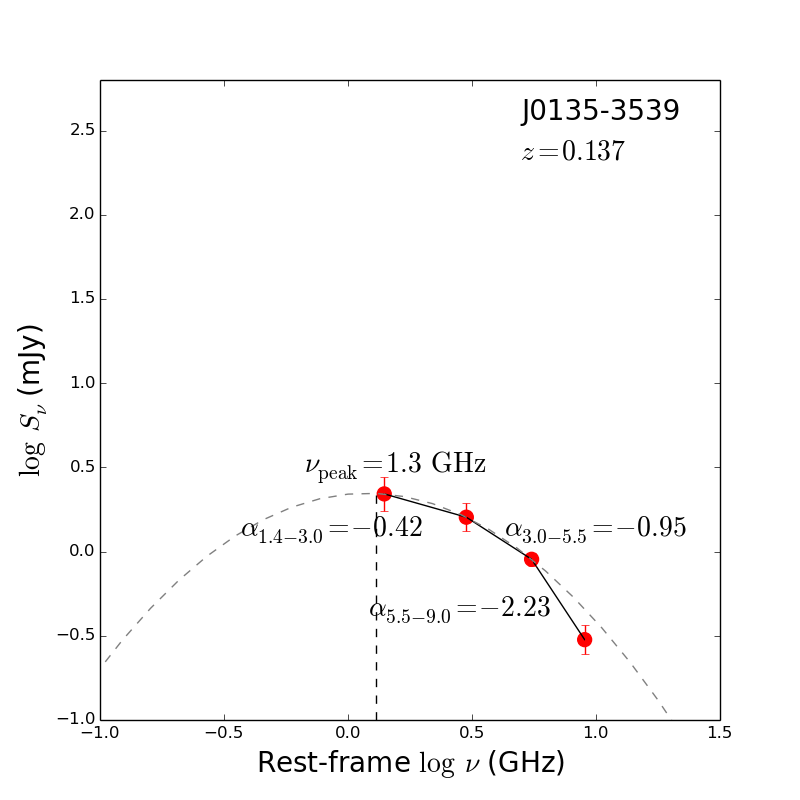}
\includegraphics[width=.43\textwidth]{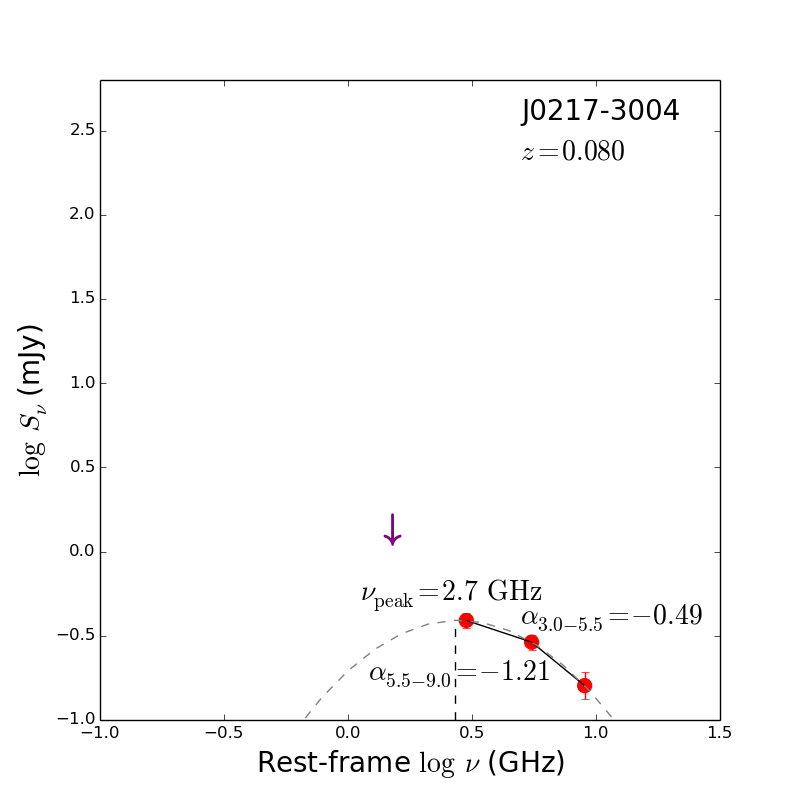}
\includegraphics[width=.43\textwidth]{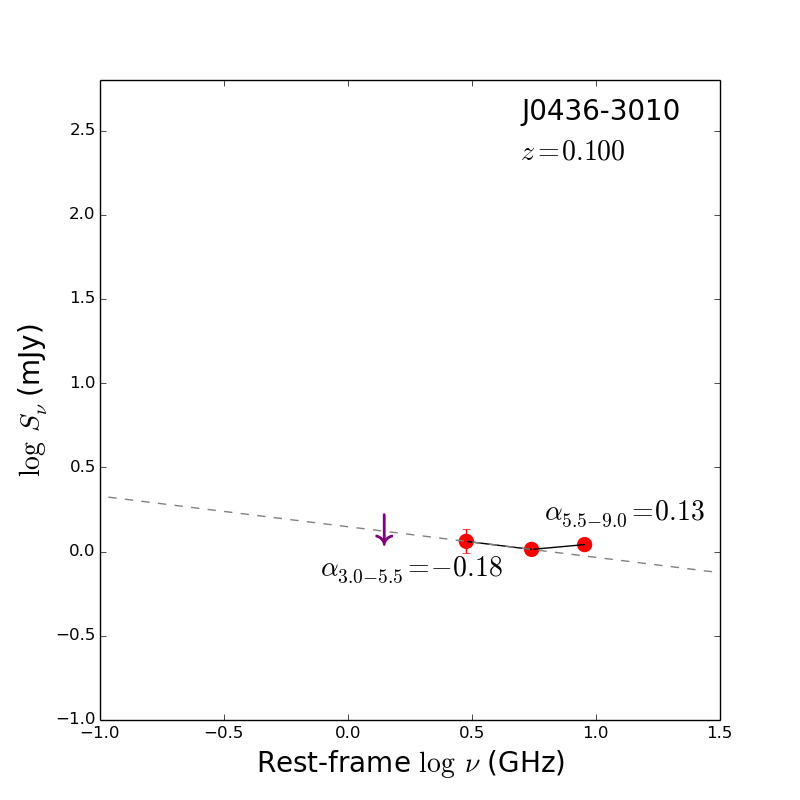}
\includegraphics[width=.43\textwidth]{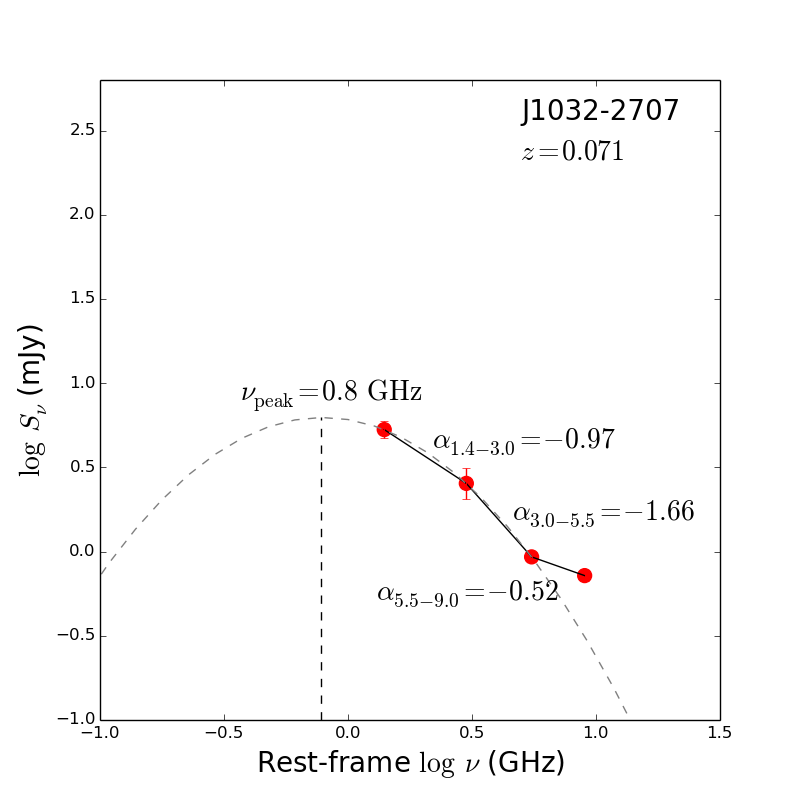}
\includegraphics[width=.43\textwidth]{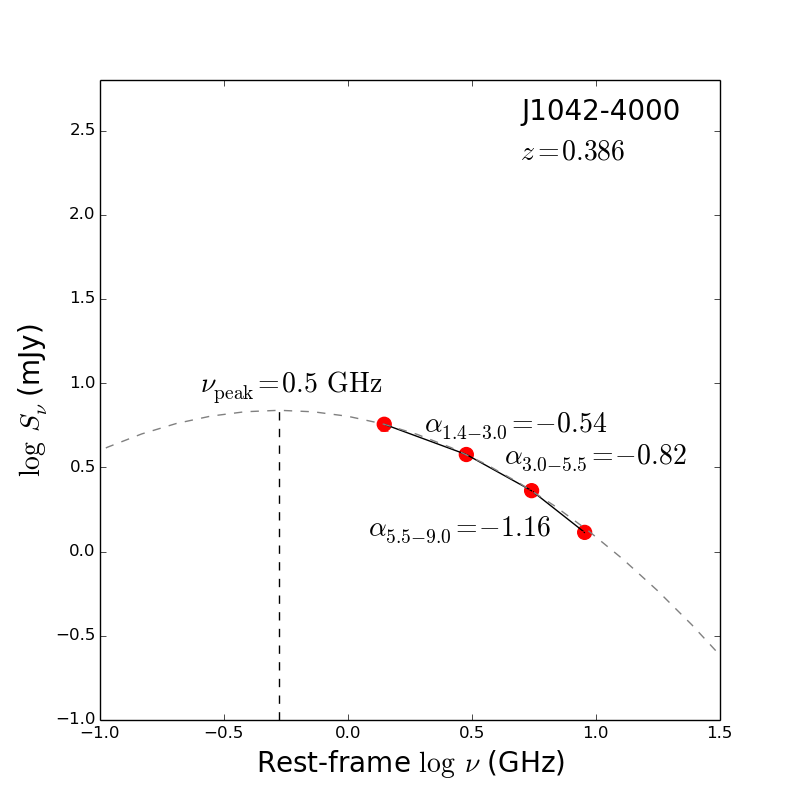}
\caption{The radio spectra of the 17 NLS1s detected in three bands or more. Detections are marked with red circles, and 3$\sigma$ upper limits are marked with purple down-arrows. The solid lines connect two detected data points. The dashed lines present the parabolic fit with three lowest-frequency data points and the turnover frequency. The object name, redshift, spectral index, and turnover frequency are labeled in each panel.}
\label{sed}
\end{figure*}

\begin{figure*}
\centering
\ContinuedFloat
\includegraphics[width=.43\textwidth]{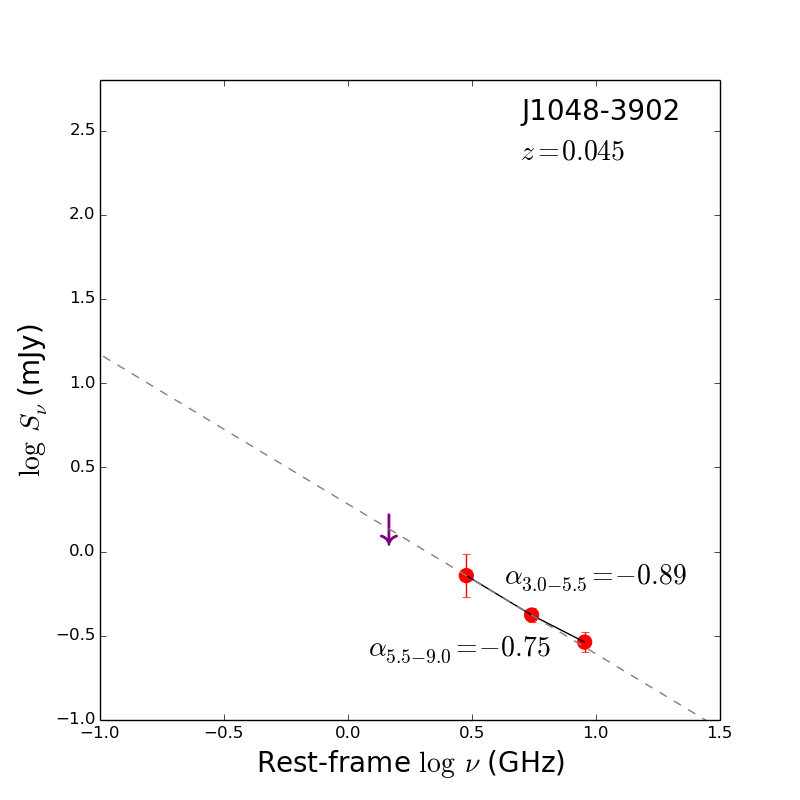}
\includegraphics[width=.43\textwidth]{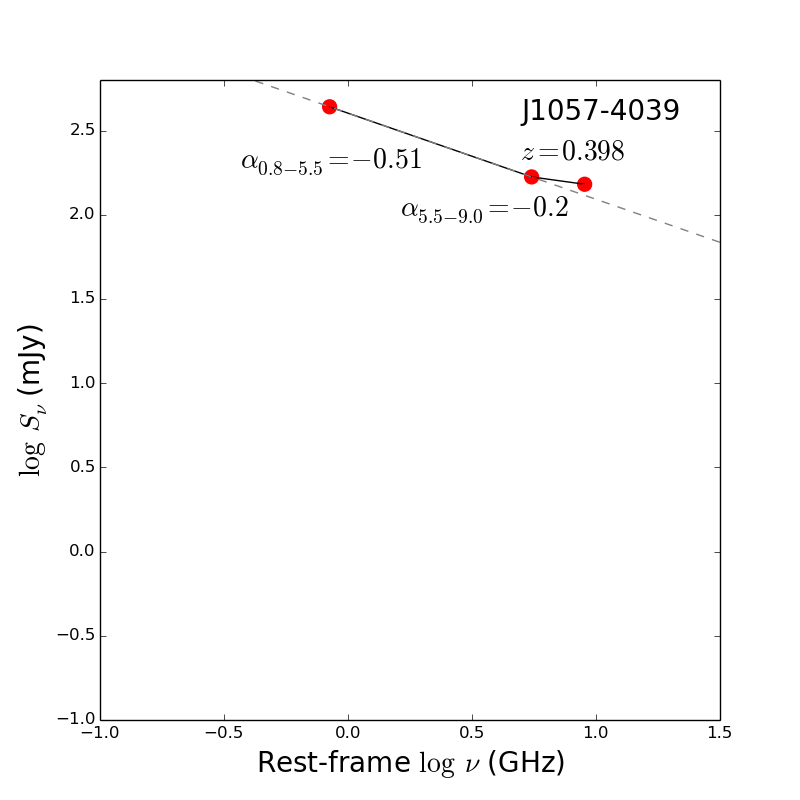}
\includegraphics[width=.43\textwidth]{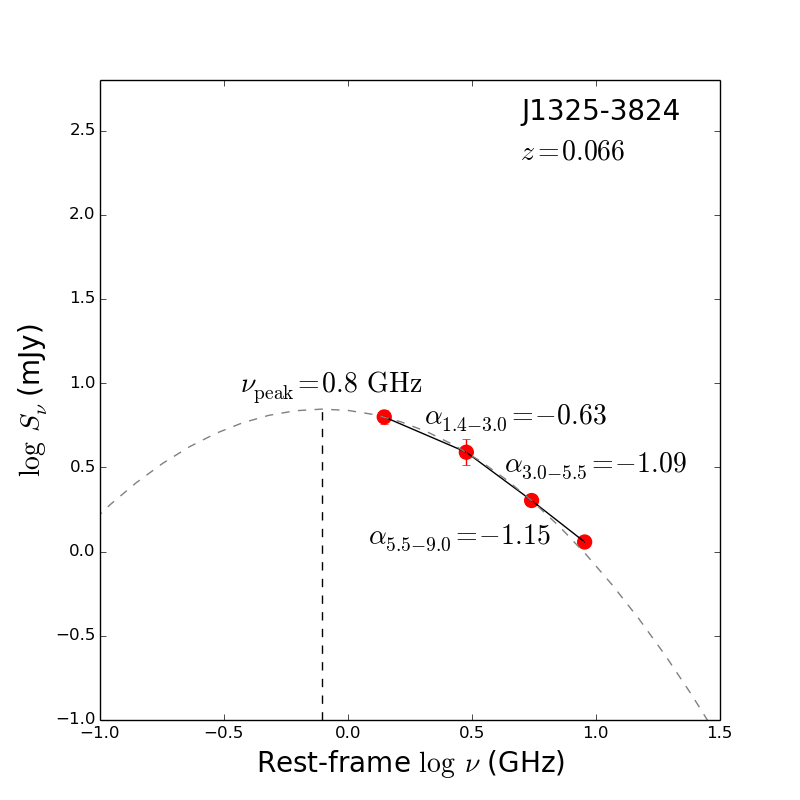}
\includegraphics[width=.43\textwidth]{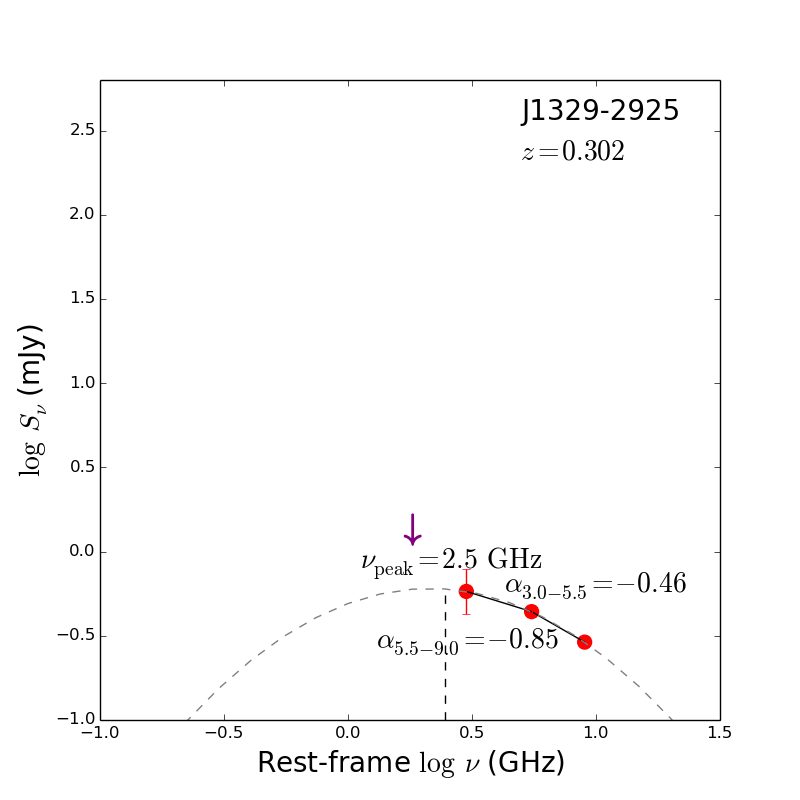}
\includegraphics[width=.43\textwidth]{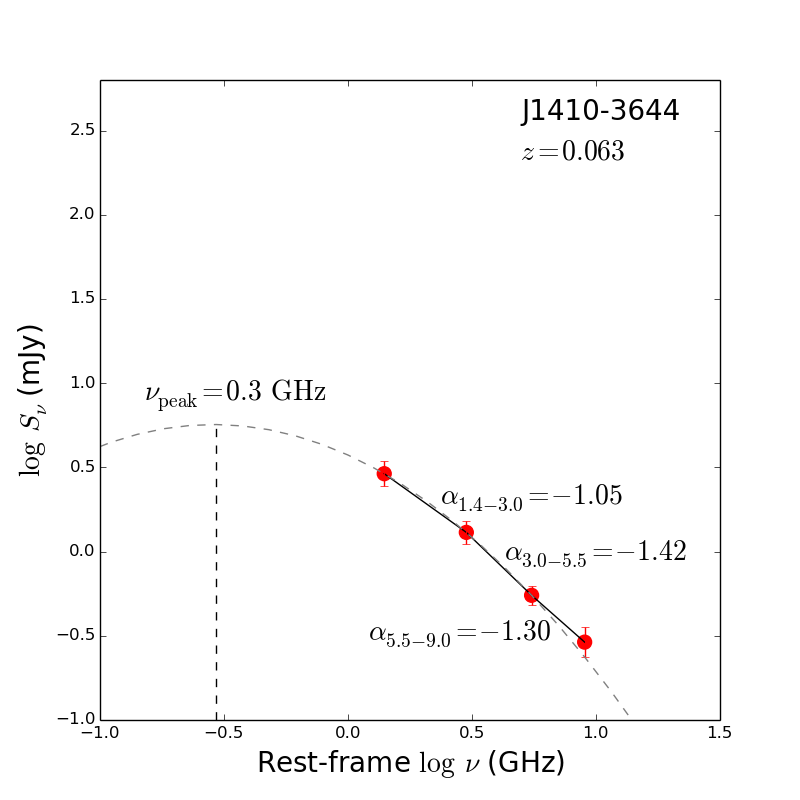}
\includegraphics[width=.43\textwidth]{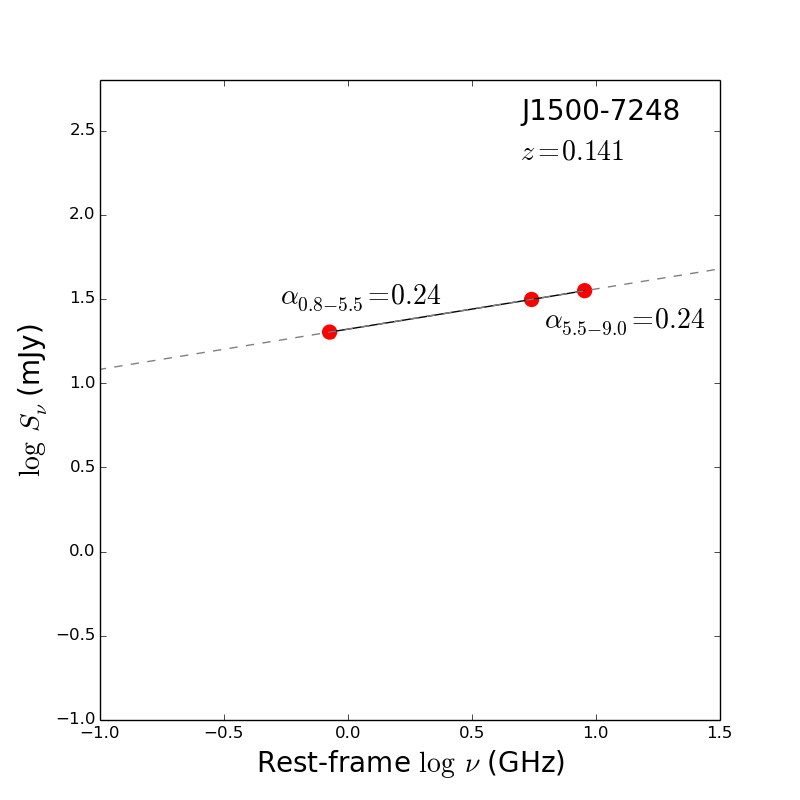}
\caption{Continued.}
\end{figure*}

\begin{figure*}
\centering
\ContinuedFloat
\includegraphics[width=.43\textwidth]{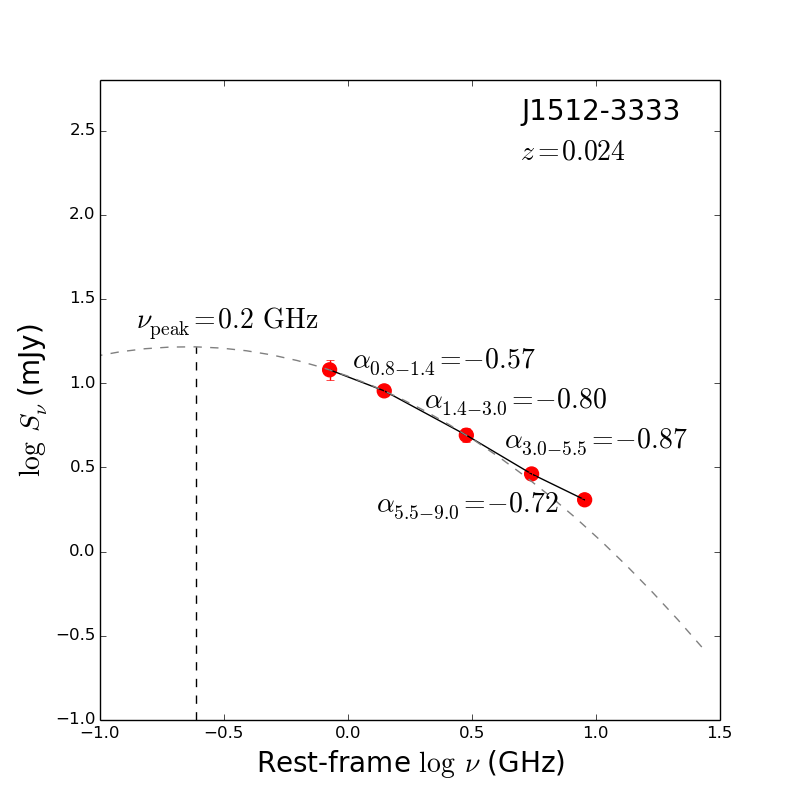}
\includegraphics[width=.43\textwidth]{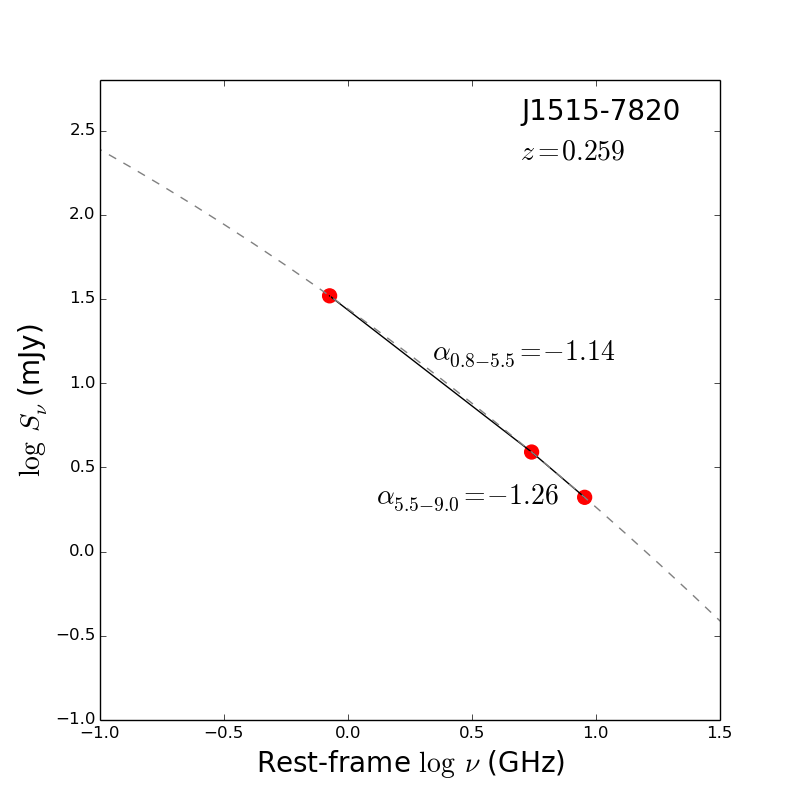}
\includegraphics[width=.43\textwidth]{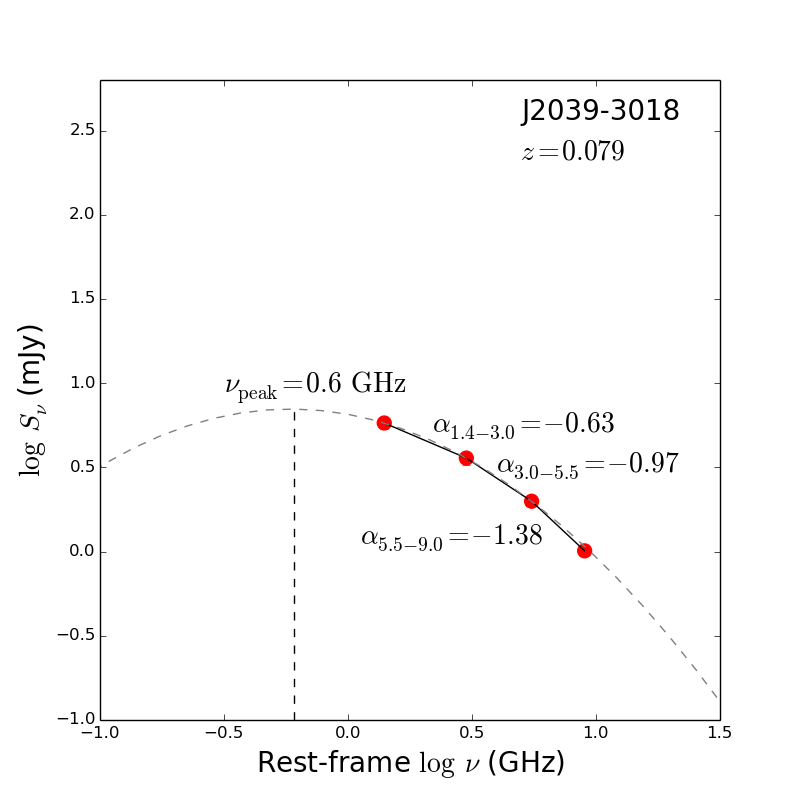}
\includegraphics[width=.43\textwidth]{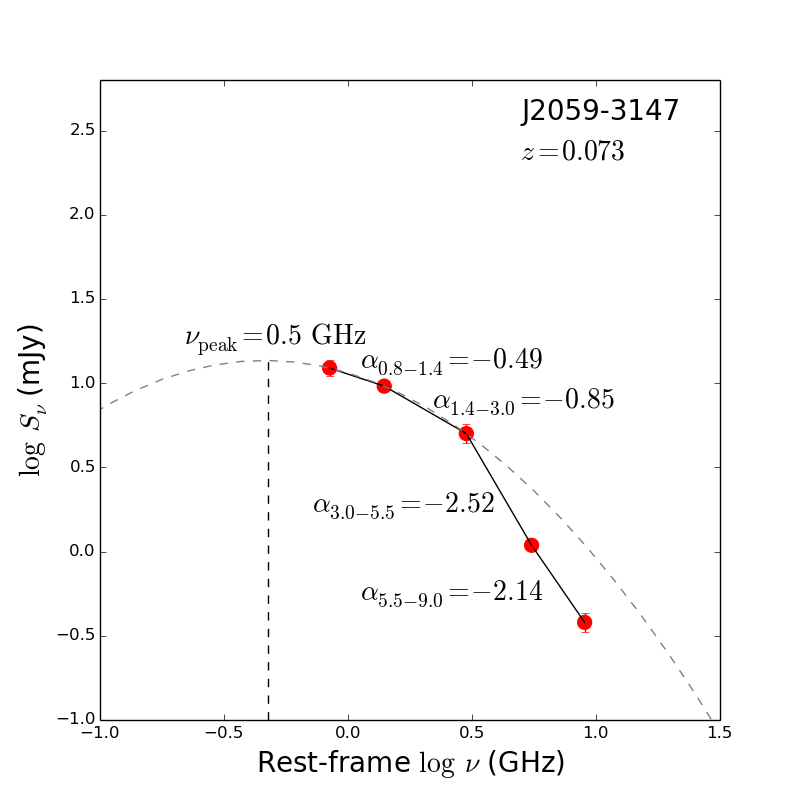}
\includegraphics[width=.43\textwidth]{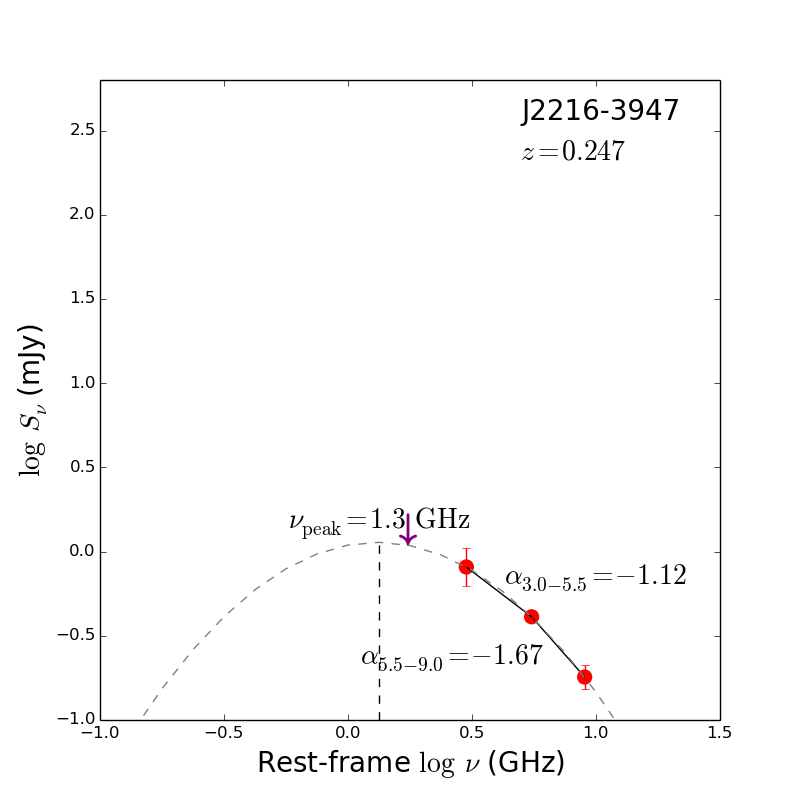}
\caption{Continued.}
\end{figure*}

There are 17 sources which are detected in three bands or more. Three of them are RL.
We model their rest frame spectra with a parabola using three lowest-frequency data points, in order to estimate their turnover frequencies $\nu_{\rm p}$ (see Fig.\ref{sed}). The estimated turnover frequencies have to be taken with caution, given the lack of low-frequency data, and should be treated as upper limits.

The observed radio spectra may be produced by a composition of various emission sources, such as SF, AGN photoionization free-free emission (FFE), AGN-driven winds/outflows, relativistic jets, and/or accretion disk coronal emission.
The SF radio emission is generally dominated by a steep synchrotron spectrum with $\alpha \simeq -0.8 \pm 0.4$ across 1--10~GHz, typically without significant spectral breaks, or with a spectral turnover at around a few hundreds MHz \citep[e.g.][]{Gim2019,An2021}.
The different spectral slopes at 1--3~GHz, 3--5~GHz, and 5--9~GHz generally seen in a given object, suggest that SF is not the dominant emission mechanism.
At an observed frequency, if the radio emission is characterized by a flat or inverted spectrum, it is dominated by a compact core, which produces optically-thick synchrotron emission, and may be associated with a core-dominated jet or an accretion disk corona.
If the radio spectrum is steep, the synchrotron emission at a given frequency is extended and optically-thin, and an AGN wind/outflow or a lobe-dominated jet may be present.
Moreover, FFE can also produce extended and flat spectral emission with $S_{\nu} \propto \nu^{-0.1}$.
But it is very faint at 1--10~GHz compared to the AGN emission, and probably not detectable in the current observations \citep[e.g.][]{Baskin2021}.

In the following analyses, we classify the different origins of the radio emission based on the radio spectral properties.
However, we should keep in mind that the observations were non-simultaneous.
The VLASS observations were carried out from 2017 to 2019, while our ATCA observations were conducted from 2018 to 2020.
The SUMSS observations were completed in 2007, and the NVSS observations were carried out in more than 20 years earlier (1993--1997).
Some of the spectral breaks may be caused by variability, if the radio source is variable and compact enough.
In addition, some uncertainties are introduced by the different angular resolutions in the different bands (ATCA 6km array: 3~arcsec at 5.5~GHz and 9.0~GHz, ATCA 750m array: 20~arcsec at 5.5~GHz and 9.0~GHz, VLASS: 2.5~arcsec at 3.0~GHz, NVSS: 45~arcsec at 1.4~GHz, SUMSS: 45~arcsec at 843~MHz).
However, since the angular resolution generally drops at lower frequencies, the aperture effect will steepen the spectra at lower frequencies, in contrast with the observed general trend of spectral flattening at lower frequencies we found below.

\subsubsection{Flat spectrum}

Three of the 17 objects (J0436$-$3010, J1057$-$4039, and J1500$-$7248) have flat or inverted spectra at 1--9~GHz, of which two are RL with $R_{\rm L}$ = 621 for J1057$-$4039 and $R_{\rm L}$ = 112 for J1500$-$7248.
Their spectra are flat with $\alpha \sim -0.51$ and $-0.20$ in J1057$-$4039, and inverted with $\alpha \sim 0.24$ in J1500$-$7248.
Hence a core-dominated relativistic jet is likely to be present in these two sources, which is consistent with their radio morphology showing a central core plus extended structures (see Fig.\ref{maps}).
A spectral turnover in either low frequencies (synchrotron self-absorption) or high frequencies (optically-thin emission), is not observed at $\nu \sim$ 1--10~GHz in either source.
In RL AGN with a misaligned jet, the turnover frequency of synchrotron emission could be a signature of the age of radio sources \citep{Fanti1995,O'Dea1998,O'Dea2021}, where older radio sources are thought to have lower turnover frequencies and be more powerful than younger ones.
If this is true, J1057$-$4039 may be more evolved than J1500$-$7248.
In contrast, J0436$-$3010 is RQ with $R_{\rm L}$ = 4 and its flat spectrum ($\alpha_{3.0-5.5} \sim -0.18$) suggests that the radio emission originates from an optically-thick compact core. The more compact emission at higher frequencies becomes more dominant with an inverted slope ($\alpha_{5.5-9.0} \sim 0.13$).

\subsubsection{Steep spectrum}

Nine sources (J0040$-$3713, J0217$-$3004, J1042$-$4000, J1325$-$3824, J1329$-$2925, J1410$-$3644, J1515$-$7820, J2039$-$3018, and J2216$-$3947) show a trend of flatter spectral slopes with decreasing frequencies.
One object, J1042$-$4000, is RL with $R_{\rm L}$ = 19, although it is close to the RL and RQ border line ($R_{\rm L}$ = 10).
J1515$-$7820 is not detected at 1.4~GHz and 3.0~GHz but detected at 843~MHz, and we assume its $\alpha_{3.0-5.5} = \alpha_{1.4-3.0} = \alpha_{0.8-5.5}$.
The median spectral indices of these nine objects show flattening toward lower frequencies, with $\alpha_{5.5-9.0} = -1.21 \pm 0.17$, $\alpha_{3.0-5.5} = -0.97 \pm 0.27$, and $\alpha_{1.4-3.0} = -0.63 \pm 0.16$ (detected in five of the nine objects).
The steep radio emission of these objects is produced by an extended source, possibly associated with an AGN wind/outflow or a lobe-dominated jet.
The spectral flattening toward low frequencies (1.4--3.0~GHz or 3.0--5.5~GHz) suggests $\nu_{\rm p} \sim$ 0.3--2.7~GHz at rest frame, excluding J1515$-$7820 where $\nu_{\rm p} < 100$~MHz.
The high-frequency (5.5--9.0~GHz) spectral slopes become steeper with $\alpha \lesssim -1$, which may indicate the aging of the relativistic electron population due to inverse-Compton, synchrotron, ionization, or bremsstrahlung energy losses \citep{Tribble1993,Komissarov1994,Murgia1999}.

\subsubsection{High frequency excess (HFE)}

The object J1032$-$2707 shows a steepening spectral slope at 1--5~GHz, from $\alpha_{1.4-3.0} \sim -0.97$ to $\alpha_{3.0-5.5} \sim -1.66$, which suggests $\nu_{\rm p}$ = 0.8~GHz at rest frame, but then flattens significantly at 5--9~GHz to $\alpha_{5.5-9.0} \sim -0.52$, which suggests that a compact core component starts to dominate.
This may be a signature of the presence of HFE observed in RQ AGN \citep{Antonucci1988,Barvainis1996}.
The HFE was reported in various studies \citep{Nagar2001,Ulvestad2001,Anderson2004,Doi2005a,Doi2016b,Baldi2021}, and it may even dominate at millimeter bands ($\sim$ 100~GHz) \citep{Doi2005b,Doi2011,Behar2015}.
The physical origin of the HFE is still unclear.
It may be produced by either self-absorbed synchrotron emission from a jet or a corona, or synchrotron or FFE which is absorbed by circumnuclear photoionized gas \citep{Laor2008,Inoue2014,Doi2016a,Raginski2016,Lahteenmaki2018,Berton2020b,Jarvela2021a,Baskin2021}.
In addition, FFE from hot or photoionized gas can not be excluded, given the large errors on the estimates of the spectral indices, which may be consistent with the expected FFE with $\alpha = -0.1$.
However, FFE of photoionized gas is expected to be rather weak producing $R_{\rm L} \sim 0.03$ at 5~GHz \citep{Baskin2021}, which is a factor of $> 10$ (see Table \ref{others}) too weak to be significant at 5~GHz in our sample.

\subsubsection{Relic emission}

In two other objects we see an opposite effect, the same steepening at low frequencies, but exceptional steepening to $\alpha < -2$ at high frequencies, rather than flattening.
Specifically, in J0135$-$3539, $\alpha_{1.4-3.0} \sim -0.42$ steepens to $\alpha_{3.0-5.5} \sim -0.95$ ($\nu_{\rm p}$ = 1.3~GHz), and becomes very steep with $\alpha_{5.5-9.0} \sim -2.23$.
In J2059$-$3147, $\alpha_{0.8-1.4} \sim -0.49$ steepens to $\alpha_{1.4-3.0} \sim -0.85$ ($\nu_{\rm p}$ = 0.5~GHz), and becomes very steep already at $\alpha_{3.0-5.5} \sim -2.52$ and remains steep at $\alpha_{5.5-9.0} \sim -2.14$.
These ultra steep drop-offs may indicate the presence of relic emission, i.e.\ significant cooling since the time of a previous AGN activity which accelerated the electrons \citep{Kharb2006,Kharb2016,Congiu2017,Congiu2020,Silpa2020}.

\subsubsection{Star formation (SF)}

The remaining two sources (J1048$-$3902 and J1512$-$3333) are characterized by uniform steep spectra with $\alpha \simeq -0.8 \pm 0.4$, without significant curvature, which may indicate that the radio emission is dominated by SF activity.
The small spectral curvature in J1512$-$3333 suggests $\nu_{\rm p}$ = 0.2~GHz at rest frame, but $\nu_{\rm p}$ can not be estimated in J1048$-$3902 due to the lack of an appropriate spectral curvature.
If the radio emission originates from SF only, which is a combination of thermal and non-thermal components \citep{Murphy2011}, the star formation rate (SFR) can be estimated based on 5~GHz luminosity via \citep{Tabatabaei2017}
\begin{equation}
\left( \frac{\rm SFR}{\rm M_{\odot} \, yr^{-1}} \right) = 4.1 \times 10^{-38} \left( \frac{\nu L_{\rm 6cm}}{\rm erg \, s^{-1}} \right)
\end{equation}
assuming a thermal electron temperature $T_{\rm e} = 10^4$~K and a spectral index of the non-thermal emission $\alpha = -1$.
The implied SFR for J1048$-$3902 and J1512$-$3333 are about 4~M$_{\odot}$~yr$^{-1}$ and 8~M$_{\odot}$~yr$^{-1}$ respectively.
We additionally note that the spectral slopes of J1325$-$3824 also reside in the range of $\alpha \simeq -0.8 \pm 0.4$, but the spectral curvature suggests $\nu_{\rm p}$ = 0.8~GHz at rest frame. If we assume the radio emission is produced by SF only, the implied SFR is about 48~M$_{\odot}$~yr$^{-1}$. The higher frequency turnover and an order of magnitude higher SFR make J1325$-$3824 different from the other two objects and disfavored the SF origin.
However, further information, for instance, the spatial distribution of the radio emission, and other SFR indicators such as various line emission, is needed to confirm the origin of the radio emission.

\subsubsection{Summary of the spectral curvature}

A summary of all the spectral curvature results of the 17 objects discussed above is presented in Fig.\ref{slope}. This figure plots the spectral indices $\alpha_{3.0-5.5}$ (or $\alpha_{0.8-5.5}$ in one RQ and two RL objects) versus $\alpha_{5.5-9.0}$, and $\alpha_{3.0-5.5}$ versus $\alpha_{1.4-3.0}$ (detected in eight of the 17 objects).
The majority, 14 of the 17 objects, reside at $\alpha < -0.5$ in all bands within 1$\sigma$, reflecting optically-thin synchrotron emission.
This trend is even stronger for the RQ NLS1s where 13 out of 14 RQ objects are steep, but is reversed for the three RL NLS1s where only one is steep.
Only three of the 17 objects, of which two are RL, reside at $\alpha > -0.5$ in both bands, indicating a very compact optically-thick synchrotron source.

In the 14 steep NLS1s, two objects are consistent with being flat above 5.5~GHz and four objects are consistent with becoming flat below 5.5~GHz within 1$\sigma$ (see Fig.\ref{slope} left panel).
Eight of these 14 objects are also detected at 1.4~GHz, and four of them are consistent with being flat below 3.0~GHz within 1$\sigma$ (see Fig.\ref{slope} right panel).
Overall, nine of the 14 steep NLS1s reside above the 1:1 ratio line of $\alpha_{3.0-5.5} = \alpha_{5.5-9.0}$, with a median spectral flattening of $\Delta \alpha = \alpha_{3.0-5.5} - \alpha_{5.5-9.0} = 0.23 \pm 0.34$ (see Fig.\ref{slope} left panel).
In eight of these 14 objects detected at 1.4~GHz, all of them reside below the 1:1 ratio line of $\alpha_{1.4-3.0} = \alpha_{3.0-5.5}$, with a median spectral flattening of $\Delta \alpha = \alpha_{1.4-3.0} - \alpha_{3.0-5.5} = 0.29 \pm 0.30$ (see Fig.\ref{slope} right panel).
This spectral flattening at lower frequencies likely reflects that the optically-thin synchrotron emission starts becoming optically-thick with decreasing frequencies, i.e.\ self-absorbed, which suggests the small size of the radio emitting region (see Section 4.2).

In the plot of $\alpha_{3.0-5.5}$ versus $\alpha_{5.5-9.0}$ (Fig.\ref{slope} left panel), three objects (J1048$-$3902, J1325$-$3824, and J1512$-$3333) lie close to the equal slope relation within the range of $\alpha = -0.8 \pm 0.4$ populated by SF galaxies, and two more objects reside within 1$\sigma$ from this area.
However in the plot of $\alpha_{3.0-5.5}$ versus $\alpha_{1.4-3.0}$ (Fig.\ref{slope} right panel), only one object (J1512$-$3333) stays in the area populated by SF galaxies (J1048$-$3902 is not detected at 1.4~GHz, and J1325$-$3824 is detected but is outside the SF area.), and three more objects reside within 1$\sigma$ from this area.

There are some outliers.
Only one object shows a clear transition from optically-thin synchrotron emission below 5.5~GHz to optically-thick above 5.5~GHz.
This HFE could become more prevalent, and possibly dominants above $\sim$ 50~GHz (see references in Section 4.1.3), and likely reflects a very compact optically-thick synchrotron source which starts dominating at higher frequencies.
Two objects show an ultra steep spectrum with $\alpha_{5.5-9.0} < -2$, and one of them is the only object with a spectrum becoming ultra steep already at $\alpha_{3.0-5.5} < -2$, which may indicate an intermittent AGN activity on time scales longer than the higher energy electron cooling times (see Section 4.2).

\begin{figure*}
\centering
\includegraphics[width=.45\textwidth]{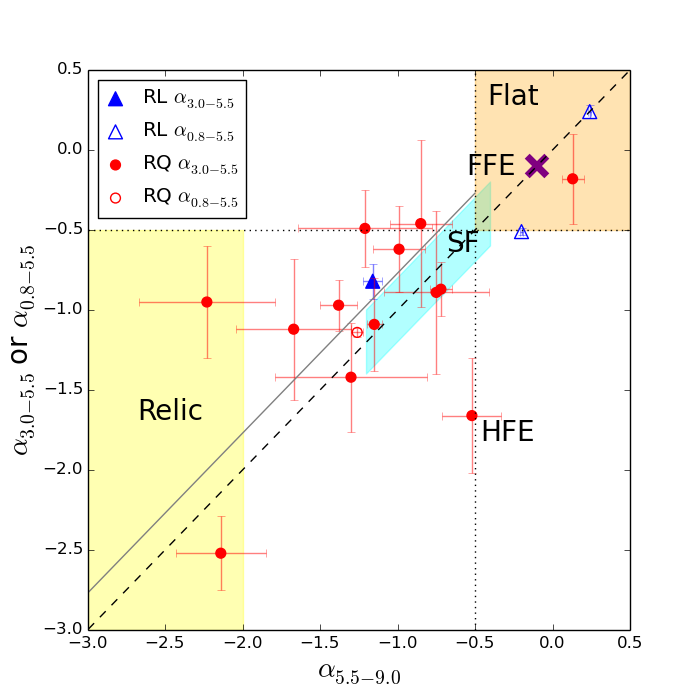}
\includegraphics[width=.45\textwidth]{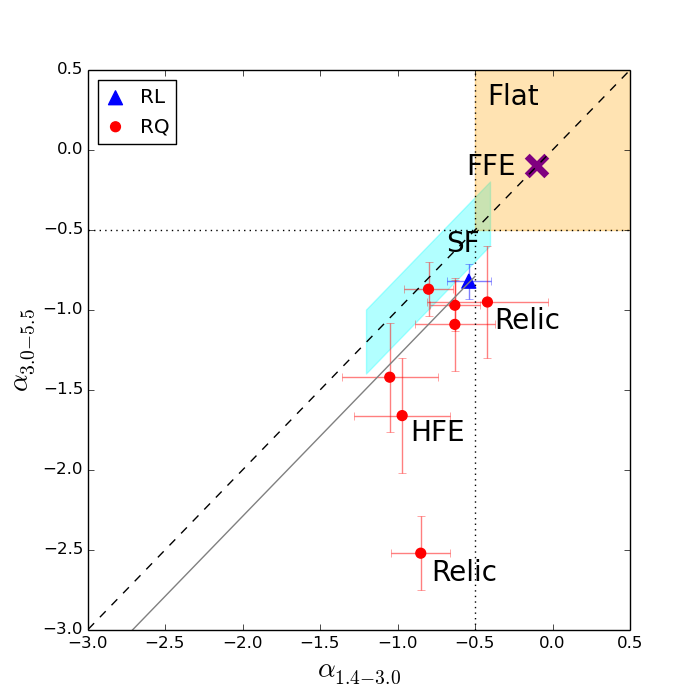}
\caption{The spectral indices $\alpha_{3.0-5.5}$ (or $\alpha_{0.8-5.5}$) versus $\alpha_{5.5-9.0}$ (left panel) and $\alpha_{3.0-5.5}$ versus $\alpha_{1.4-3.0}$ (right panel).
The blue triangles and red circles represent RL and RQ objects.
The filled and empty symbols in the left panel show $\alpha_{3.0-5.5}$ and $\alpha_{0.8-5.5}$ respectively.
The black dashed line is the 1:1 ratio line of equal slopes.
The horizontal and vertical black dotted lines are $\alpha$ = $-$0.5.
The area showing flat or inverted spectra, which may be associated with a core-dominated jet or a corona, is filled in orange.
The area with a spectral index $\alpha = -0.8 \pm 0.4$ with a scatter of 0.2, which may be related to SF, is filled in cyan.
The area showing very steep spectral slopes with $\alpha_{5.5-9.0} < -2$ in the left panel, which may imply possible relic emission, is filled in yellow.
The purple cross places in where FFE is produced.
An object exhibiting HFE is also labeled.
The grey lines of $\Delta \alpha = \alpha_{3.0-5.5} - \alpha_{5.5-9.0} = 0.23$ in the left panel, and of $\Delta \alpha = \alpha_{1.4-3.0} - \alpha_{3.0-5.5} = 0.29$ in the right panel, are plotted to guide the eyes.}
\label{slope}
\end{figure*}

\subsubsection{Additional sources}

\begin{figure*}
\centering
\includegraphics[width=.43\textwidth]{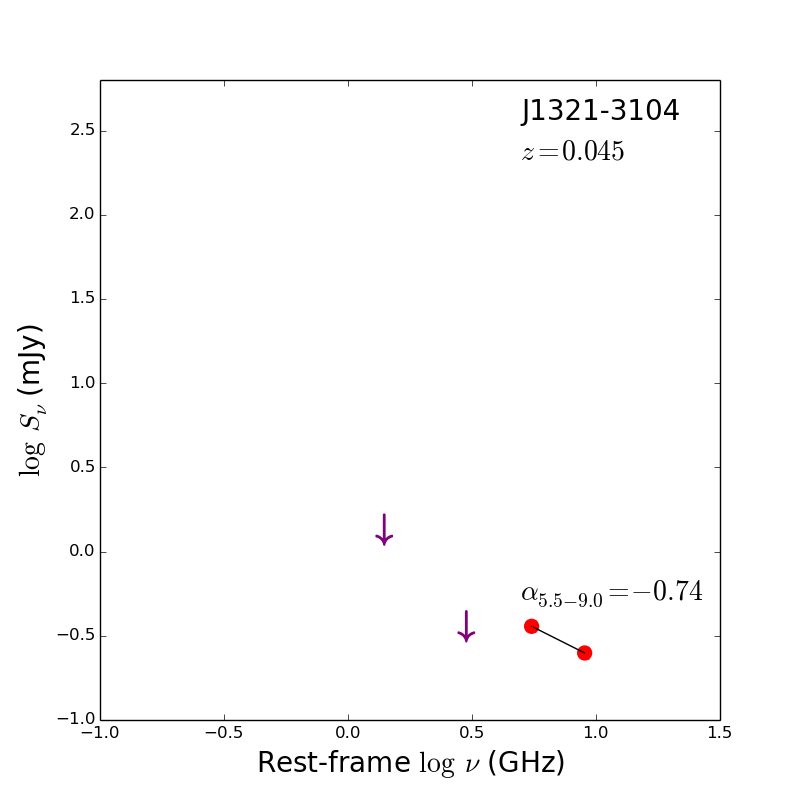}
\includegraphics[width=.43\textwidth]{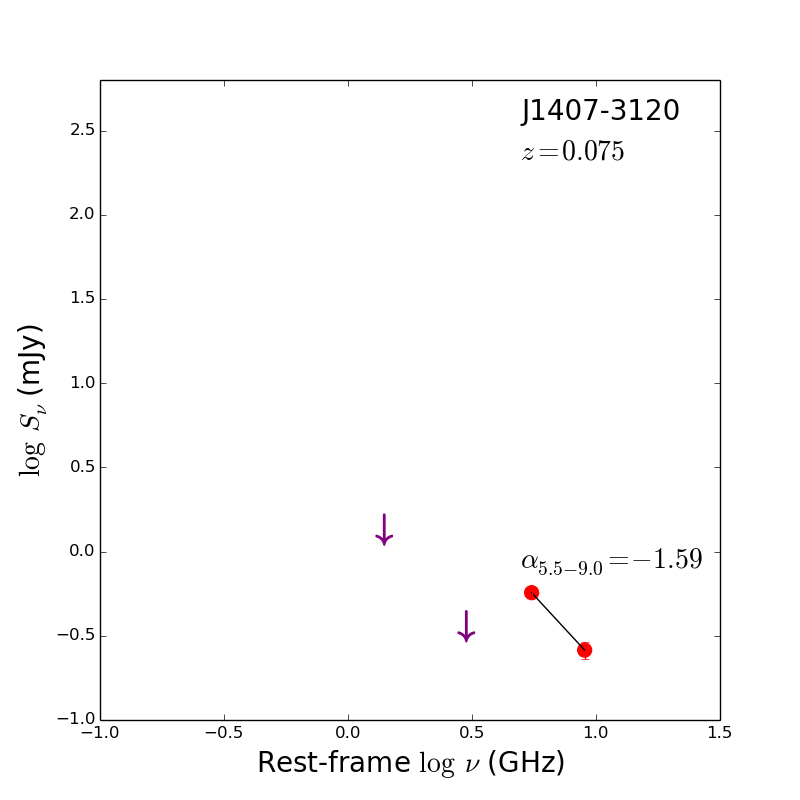}
\caption{The radio spectra of the two NLS1s where the upper limits suggest a peak frequency $\sim$ 6~GHz at rest frame. Detections are marked with red circles, and 3$\sigma$ upper limits are marked with purple down-arrows. The solid lines connect two detected data points. The object name, redshift, and spectral index are labeled in each panel.}
\label{peak5GHz}
\end{figure*}

The remaining 25 sources, of the 42 objects detected, are detected in either one (10 objects) or two (15 objects) bands.
We do not have further details of their intrinsic spectral curvature.
However, the additional 15 values of $\alpha_{5.5-9.0}$ are again mostly steep with 12 having $\alpha_{5.5-9.0} < -0.5$, and the three flat objects have $\alpha_{5.5-9.0} > -0.5$ at less than 1$\sigma$.
In total, of the 32 NLS1s detected at both 5.5~GHz and 9.0~GHz, 26 objects show steep spectra, where only one source is RL, and six objects have flat spectra, where two sources are RL.
Two objects, J1321$-$3104 and J1407$-$3120, show a steep spectrum at $\alpha_{5.5-9.0} < -0.5$, but the upper limits suggest an inverted spectrum at $\alpha_{3.0-5.5} > 0$, as shown in Fig.\ref{peak5GHz}, which implies $\nu_{\rm p} \sim$ 6~GHz at rest frame.
Also, some objects have extremely flat or steep slopes with $\alpha_{5.5-9.0} \gtrsim 2$ or $\lesssim -3$. These extreme values have to be taken with caution because of possible systematic errors caused by the low S/N data in one of the bands.

\subsection{Constraints on the sizes and the cooling time scales}

The synchrotron self-absorption frequency, as estimated based on $\nu_{\rm p}$, can be used to constrain the size of the radio emitting region.
We make these estimates for 14 objects, which include 12 objects where $\nu_{\rm p}$ is derived from the parabolic fit, and two objects with upper limits which suggest $\nu_{\rm p} \sim$ 6~GHz.
We also need an estimate of the strength of the magnetic field $B$, which we derive by assuming that the magnetic energy density is in equipartition with the photon energy density.
The implied size of the synchrotron source is \citep[Eq.22 in][]{Laor2008}
\begin{equation}
R = 0.47 L_{30}^{0.4} L_{46}^{0.1} \nu_{\rm p}^{-1}
\end{equation}
where $R$ is the radius of the radio-sphere in pc, $\nu_{\rm p}$ is the turnover frequency in GHz, $L_{30}$ is the luminosity density at the turnover frequency in 10$^{30}$~erg~s$^{-1}$~Hz$^{-1}$, and $L_{46}$ is the bolometric luminosity in 10$^{46}$~erg~s$^{-1}$.
The shortest variability time scale is provided by the light crossing time of the radio-sphere, $t_{\rm light} = R / c$, if the emission is not relativistically beamed.

The value of $B$, as estimated above based on equipartition, is given by \citep[Eq.21 in][]{Laor2008}
\begin{equation}
B = 0.27 R^{-1} L_{46}^{1/2}
\end{equation}
where $B$ is the magnetic field in Gauss.
The $\gamma$ factor of the electrons which produce synchrotron emission at a given frequency of $\nu_{\rm obs}$ in GHz can then be computed using \citep[Eq.30 in][]{Laor2008}
\begin{equation}
\nu_{\rm obs} = \gamma^2 B / 820.
\end{equation}
Assuming $\nu_{\rm obs}$ = 5~GHz, the time scale for the electrons at 5~GHz to lose half of its energy through synchrotron emission can be evaluated via \citep[Eq.29 in][]{Laor2008}
\begin{equation}
t_{\rm syn} = 5.1 \times 10^8 \gamma^{-1} B^{-2}
\end{equation}
where $t_{\rm syn}$ is the synchrotron cooling time at a given frequency in s.

\setcounter{table}{5}

\begin{table}
\centering
\caption{The estimates of the synchrotron source size, the light crossing time, the magnetic field, and the synchrotron cooling time.}
\begin{footnotesize}
\begin{tabular}{ccccccc}
\hline
\hline
\multirow{2}{*}{Name} & $\nu_{\rm p}$ & $S_{\nu}$ & $R$ & $t_{\rm light}$ & $B$ & $t_{\rm syn}$ \\
& (GHz) & (mJy) & (pc) & (year) & (Gauss) & (year) \\
(1) & (2) & (3) & (4) & (5) & (6) & (7) \\
\hline
J0040$-$3713 & 1.6 & 1.2 & 0.05 & 0.15 & 0.46 & 0.80 \\
J0135$-$3539 & 1.3 & 2.2 & 0.30 & 0.98 & 0.29 & 1.58 \\
J0217$-$3004 & 2.7 & 0.4 & 0.04 & 0.12 & 0.69 & 0.44 \\
J1032$-$2707 & 0.8 & 6.3 & 0.39 & 1.27 & 0.15 & 4.47 \\
J1042$-$4000 & 0.5 & 6.9 & 3.74 & 12.20 & 0.06 & 18.92 \\
J1325$-$3824 & 0.8 & 7.0 & 0.45 & 1.47 & 0.30 & 1.58 \\
J1329$-$2925 & 2.5 & 0.6 & 0.21 & 0.68 & 0.69 & 0.44 \\
J1410$-$3644 & 0.3 & 5.7 & 0.94 & 3.08 & 0.08 & 11.36 \\
J1512$-$3333 & 0.2 & 16.4 & 0.73 & 2.36 & 0.02 & 66.09 \\
J2039$-$3018 & 0.6 & 7.0 & 0.55 & 1.79 & 0.07 & 13.05 \\
J2059$-$3147 & 0.5 & 13.6 & 0.86 & 2.81 & 0.06 & 15.79 \\
J2216$-$3947 & 1.3 & 1.1 & 0.41 & 1.32 & 0.29 & 1.66 \\
\hline
J1321$-$3104 & 5.7 & 0.4 & 0.01 & 0.03 & 2.59 & 0.06 \\
J1407$-$3120 & 5.9 & 0.6 & 0.02 & 0.06 & 1.80 & 0.10 \\
\hline
Median & 1.05 & 3.95 & 0.40 & 1.30 & 0.29 & 1.62 \\
\hline
\end{tabular}
\end{footnotesize}
\label{turnover}
\flushleft{\textbf{Notes.} Columns: (1) name, (2) turnover frequency at rest frame, (3) flux density at turnover frequency, (4) radius of the radio-sphere, (5) light crossing time, (6) magnetic field, (7) synchrotron cooling time at 5~GHz.
The peak frequency of the first 12 objects is derived from the parabolic fit, and the peak frequency of the last 2 objects is suggested by upper limits.
The line at the end of the table shows the median values of the different parameters.}
\end{table}

The estimates of the synchrotron source size, the light crossing time, the magnetic field, and the synchrotron cooling time are reported in Table \ref{turnover}.
All RQ sources are very compact with $R \sim$ a fraction of 1~pc.
In the only one RL source (J1042$-$4000), $R \sim$~3.7~pc, an order of magnitude larger.
In the two objects with $\nu_{\rm p} \sim$~6~GHz, we get $R \sim$~0.01--0.02~pc, an order of magnitude smaller.
The magnetic field is also typically a fraction of Gauss, and ranges in strength of 0.02--2.59~Gauss.

The synchrotron cooling time for the electrons at 5~GHz varies from 0.06 year (3 weeks) to 66 years, and is typically longer than or comparable to the light crossing time.
The electrons can also cool via Compton scattering, but since we assume equipartition of the radiation and the magnetic energy density, the synchrotron cooling time is by assumption equal to the Compton cooling time.
Adiabatic cooling will be at the sound crossing time, which is significantly longer than the light crossing time, and therefore not likely to be much shorter than the synchrotron/Compton cooling time.
Only the Coulomb collision cooling time scale can be significantly shorter than $t_{\rm syn}$ if the relativistic electrons interact with rather dense gas with a number density of $> 10^7$~cm$^{-3}$ \citep[Eq.37 in][]{Laor2008}, as is likely prevalent in the accretion disk surface.
However, the expected variability time scale can not be shorter than the light crossing time scale, and may be longer or comparable if synchrotron cooling is the dominant cooling mechanism.

Since all cooling time scales become shorter at higher frequencies, we may have an intermediate situation where the electrons above a certain energy had time to cool since their last acceleration, while the electrons below that energy still maintain their initial energy.
This may explain the preferential steepening to $\alpha < -2$, or the so-called relic emission, observed at higher frequencies in J0135$-$3539 and J2059$-$3147.
If this is the case, the radio emitting electrons at 5~GHz in J0135$-$3539 were produced in an acceleration event $\sim$ 1.6 years ago.
In J2059$-$3147, $t_{\rm syn} \sim$ 15.8 years at 5~GHz, but since the steepening in this object already starts at 3~GHz, where $\gamma$ is somewhat smaller, the acceleration event might occur $\sim$ 22.2 years ago.

The two objects (J1321$-$3104 and J1407$-$3120) with $\nu_{\rm p} \sim$ 6~GHz have the most compact core (0.01--0.02~pc), the highest magnetic field (1.8--2.6~Gauss), and the shortest $t_{\rm syn}$ of a month time scale.
In contrast, J1512$-$3333 which may be dominated by SF activity based on its spectral shape, has the lowest turnover frequency (0.2~GHz), the lowest magnetic field (0.02~Gauss), and the longest variability time scale (several decades).

The estimates of the various variability time scales above are not robust, as the assumed equipartition used to derive $B$ may not hold. A detection of radio variability can therefore provide a direct constraint on the possible maximal size of the synchrotron emitting region and on the electron cooling mechanisms.

\subsection{Multi-property correlations}

The correlation between the radio spectral slopes at 5--9~GHz and the Eddington ratios was studied by \citet{Laor2019} with a sample of 25 RQ Palomar-Green (PG) quasars.
They found that RQ quasars with high Eddington ratios ($L_{\rm bol}/L_{\rm Edd} > 0.3$) show steep ($\alpha < -0.5$) optically-thin synchrotron emission, which is produced by an extended source, possibly a weak jet or wind component.
In contrast, low Eddington ratios ($L_{\rm bol}/L_{\rm Edd} < 0.3$) RQ quasars show flat ($\alpha > -0.5$) optically-thick synchrotron emission, which is produced by a compact core, possibly a weak jet base or an accretion disk corona.

\begin{figure}
\centering
\includegraphics[width=.5\textwidth]{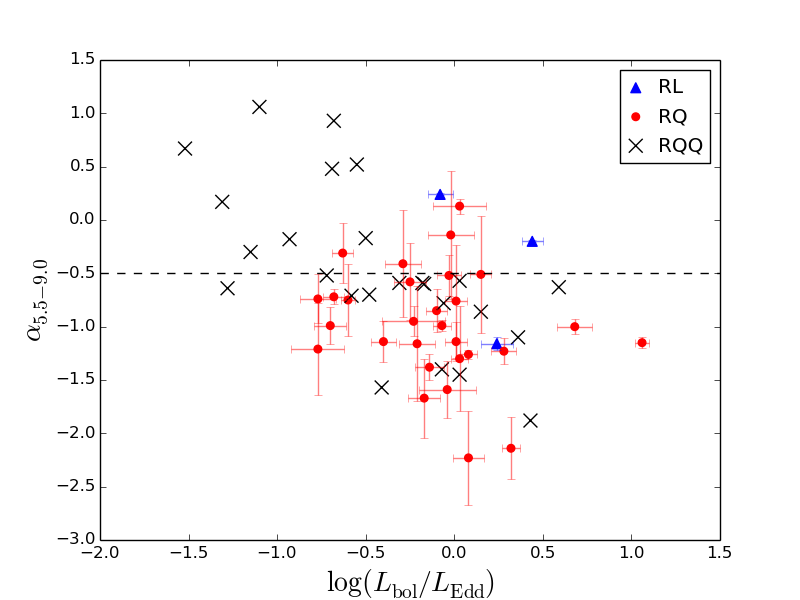}
\includegraphics[width=.5\textwidth]{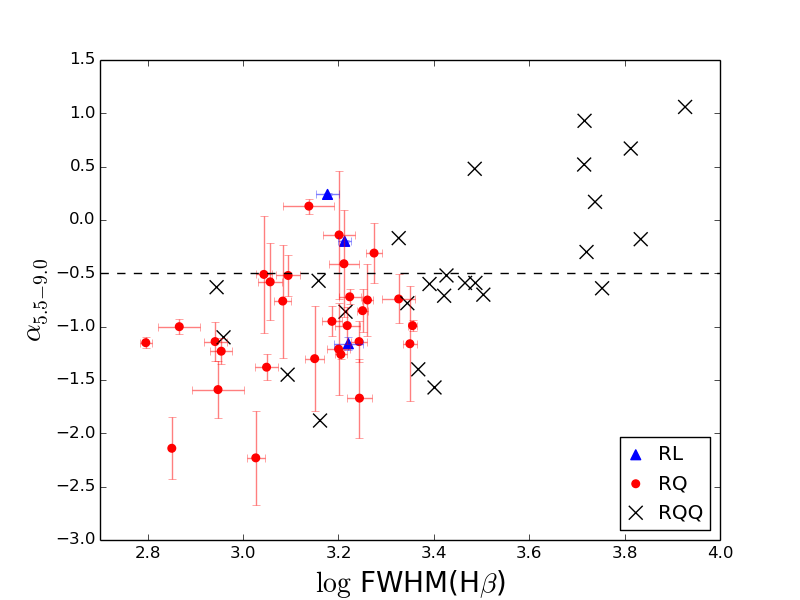}
\includegraphics[width=.5\textwidth]{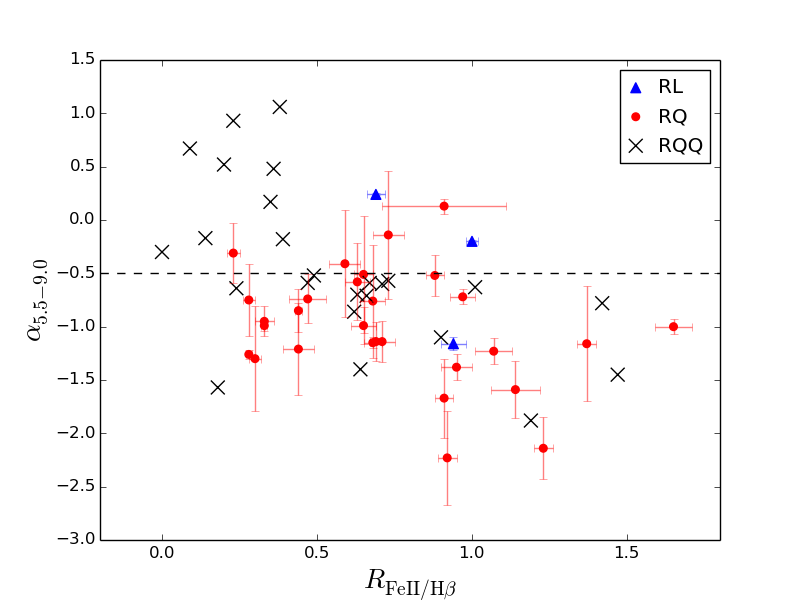}
\caption{
{\it Top panel:} The spectral index $\alpha_{5.5-9.0}$ vs.\ the Eddington ratio.
{\it Middle panel:} The spectral index $\alpha_{5.5-9.0}$ vs.\ the FWHM of H$\beta$ line.
{\it Bottom panel:} The spectral index $\alpha_{5.5-9.0}$ vs.\ the flux ratio of Fe~II / H$\beta$.
Red circles and blue triangles represent RQ and RL NLS1s respectively. Black crosses represent RQ quasars from \citet{Laor2019}.}
\label{correlation}
\end{figure}

Fig.\ref{correlation} presents the distributions of $\alpha_{5.5-9.0}$ against the Eddington ratio, the FWHM of H$\beta$ line, and the flux ratio of Fe~II / H$\beta$ for the 32 NLS1s in our sample which have $\alpha_{5.5-9.0}$ measurements.
The other 10 objects which only have upper and lower limits are excluded as these limits are uncertain.
We note that the radio emission processes of RL and RQ sources are different.
But we include the RL objects, as well as the RQ quasars from \citet{Laor2019}, in the plots for comparison.
The NLS1s in our sample, which are selected to have a low H$\beta$ FWHM, are characterized by a high $L_{\rm bol}/L_{\rm Edd}$ as expected, and tend to have a high $R_{\rm{Fe~II}/\rm{H}\beta}$, which are also features of NLS1s \citep{Boroson1992,Yang2020}.
Their distributions overlap the distributions of the 25 RQ quasars, which extend to broader H$\beta$ FWHM, and lower $L_{\rm bol}/L_{\rm Edd}$ and $R_{\rm{Fe~II}/\rm{H}\beta}$ values.
This overlap is expected since a significant fraction of the PG quasars are in fact NLS1s.

\begin{table*}
\centering
\caption{The Spearman’s rank-order correlations between radio spectral slope $\alpha_{5.5-9.0}$ and Eddington ratio, FWHM of H$\beta$ line, and flux ratio of Fe~II / H$\beta$.}
\begin{footnotesize}
\begin{tabular}{ccccccc}
\hline
\hline
\multirow{2}{*}{Sample} & \multicolumn{2}{c}{$\log (L_{\rm bol}/L_{\rm Edd})$} & \multicolumn{2}{c}{$\log \rm{FWHM}(\rm{H}\beta)$} & \multicolumn{2}{c}{$R_{\rm{Fe~II}/\rm{H}\beta}$} \\
& $r_s$ & $p$ & $r_s$ & $p$ & $r_s$ & $p$ \\
\hline
RQ NLS1s & $-$0.29 & 0.12 & 0.31 & 0.10 & $-$0.32 & 0.093 \\
RQ quasars & $-$0.69 & 1.3 $\times$ 10$^{-4}$ & 0.67 & 2.6 $\times$ 10$^{-4}$ & $-$0.62 & 1.0 $\times$ 10$^{-3}$ \\
All RQ & $-$0.52 & 5.9 $\times$ 10$^{-5}$ & 0.54 & 2.2 $\times$ 10$^{-5}$ & $-$0.48 & 2.8 $\times$ 10$^{-4}$ \\
\hline
\end{tabular}
\end{footnotesize}
\label{spearmanr}
\end{table*}

We further perform the Spearman’s rank-order correlations between $\alpha_{5.5-9.0}$ and $\log (L_{\rm bol}/L_{\rm Edd})$, $\log \rm{FWHM}(\rm{H}\beta)$, and $R_{\rm{Fe~II}/\rm{H}\beta}$ for the RQ NLS1s, the RQ quasars, and the total RQ sources. The coefficient ($r_s$) and significance level ($p$) are reported in Table \ref{spearmanr}.
We do not find any significant correlation in the RQ NLS1 sample only due to the selection of high $L_{\rm bol}/L_{\rm Edd}$ objects only.
In contrast, the RQ quasars show strong and significant correlations, since they extend over a wider range of these three parameters explored \citep{Laor2019}.
The combination of the two samples leads to higher $r_s$ values than those found for the NLS1s only, but lower $r_s$ values than those found for the RQ quasars only.
The significance of the correlations in the combined sample increases compared to the RQ quasars only, as the combined sample size of 57 RQ objects is significantly larger than the 25 RQ quasars sample size.

The FWHM of H$\beta$ line and the flux ratio of Fe~II / H$\beta$ are part of the Eigenvector 1 (EV1) correlations, where NLS1s are typically high EV1 objects \citep{Boroson1992}.
The main physical driver of EV1 has long been suspected to be the Eddington ratio \citep{Boroson1992,Boroson2002,Shen2014}.
The physical mechanism behind this could be an outflow driven by radiation pressure.
The strength of the outflow depends on the ratio of radiation pressure to gravity as indicated by the Eddington ratio.
A high Eddington ratio, which is related to AGN fueling, may be coupled with a high metallicity, which is connected to starbursts activity \citep{Shemmer2002,Shemmer2004}, though some studies found no trend between Eddington ratio and metallicity \citep[e.g.][]{Warner2004}.
Such an outflow may interact with the ambient medium and generate shocks that accelerate particles to relativistic energies \citep{Jiang2010}.
These relativistic particles move in the shocked gas magnetic field resulting in optically-thin synchrotron emission \citep{Zakamska2016}.
If this scenario is true, the radio emission of the majority of NLS1s in our sample may be dominated by an AGN outflow.
Indeed, NLS1s characterized by high $L_{\rm bol}/L_{\rm Edd}$ are likely to power an outflow by means of the radiation pressure coming from the accretion disk \citep{Proga2000,Proga2007,Tombesi2010,Parker2017}.

If the constraint on the size of the radio emitting region, as derived based on the estimated $\nu_{\rm p}$, is valid then the wind has to be confined to pc scales in most objects, that is just outside the BLR.
This compact size clearly excludes SF as a possible mechanism, which has to extend at least to the galaxy core size, on sub-kpc scales.
An alternative radio source is a compact weak jet.
However, a fraction of RQ AGN with low Eddington ratios, e.g.\ low-luminosity AGN, also have the capability of producing jets \citep{Nagar2002,Gallimore2006,Kharb2021}.
Thus the weak jet found in low $L_{\rm bol}/L_{\rm Edd}$ objects raises a question why optically-thin radio emission favors to be observed at high $L_{\rm bol}/L_{\rm Edd}$ objects (e.g.\ Fig.\ref{correlation} top panel).
Further studies are necessary to clarify these issues.

\section{Summary}

In this work, we present the results of the ATCA observations of 85 NLS1s, drawn from a sample of 168 NLS1s in the southern hemisphere, with 6km and 750m arrays at 5.5~GHz and 9.0~GHz.
The NLS1s are characterized by $L_{\rm bol} / L_{\rm Edd} > 0.15$.
About half of the sample (42 objects) are detected, of which 32 are detected in both bands.
Combinations of the available archival data from VLASS, NVSS, and SUMSS at 3.0~GHz, 1.4~GHz, and 0.843~GHz respectively, allow to explore the spectral curvatures using at least three bands in 17 NLS1s (14 RQ and 3 RL).
We explore the distribution of the spectral shapes, and the correlations between the radio spectral slope and the optical properties.
The results are summarized as follow.
We note that some uncertainties are introduced due to the non-simultaneous observations and the different angular resolutions.

1. The typical radio spectral slope at 5.5--9.0~GHz is steep, with a median value of $\alpha_{5.5-9.0} = -0.99 \pm 0.20$, indicating optically-thin synchrotron emission.

2. The spectral slope typically flattens at lower frequencies, with a median value at 1.4--3.0~GHz of $\alpha_{1.4-3.0} = -0.72 \pm 0.23$.
A parabolic fit to three lowest frequencies in 12 of the 17 NLS1s suggests a median spectral turnover of $\sim$~1~GHz.
Such a turnover, if due to synchrotron self-absorption, implies the radio source smaller than a fraction of 1~pc.
The radio emission is then associated with a compact optically-thin source, possibly a compact wind/outflow or a weak jet.

3. Three objects, two of which are RL, show a flat or inverted spectrum, indicating an optically-thick source, likely a relativistic jet in the two RL NLS1s, and a weak jet base or a corona in the one RQ NLS1.

4. In two objects, the spectral slope steepens significantly to $\alpha < -2$ above 3~GHz or 5~GHz, which may suggest relic emission from an earlier intermittent AGN activity.
Based on the synchrotron cooling time, this activity occurred about 2 years ago in one object, and about 22 years ago in the other object.

5. In one object, the slope steepens from $-0.97$ to $-1.66$ at 1--5~GHz, but becomes flatter to $-0.52$ at 5--9~GHz.
This may be analogous to the HFE emission which is a common feature in RQ AGN above $\sim$ 50~GHz, where a compact optically-thick source starts dominating the emission.

6. Only two sources show a rather uniform steep slope, which may manifest radio emission from SF activity.
The implied SFR are about 4~M$_{\odot}$~yr$^{-1}$ and 8~M$_{\odot}$~yr$^{-1}$.

7. The radio slopes $\alpha_{5.5-9.0}$ are measured in 32 NLS1s, of which three are RL.
The slopes are steep in 25 of the 29 RQ objects, which is consistent with earlier studies of steep radio emission in RQ quasars with $L_{\rm bol}/L_{\rm Edd} > 0.3$.
The distributions of the radio slope versus the H$\beta$ FWHM, the Eddington ratio, and the Fe~II/H$\beta$ flux ratio, overlap the distributions found earlier for RQ quasars.

Finally, we highlight that repeated simultaneous radio observations over a wide frequency range, and over a range of resolutions, carry crucial information on various processes in AGN on different scales, and may also include a record of the recent history of the AGN activity.

\section*{Acknowledgements}

We thank the anonymous referee for suggestions leading to the improvement of this work.
A.L. acknowledges support by the Israel Science Foundation (grant No.1008/18).
M.F.G. is supported by the National Science Foundation of China (grant No.11873073), Shanghai Pilot Program for Basic Research - Chinese Academy of Science, Shanghai Branch (JCYJ-SHFY-2021-013), and the science research grants from the China Manned Space Project (CMSCSST-2021-A06).
M.B. is an ESO fellow.
E.J. is a current ESA research fellow.
E.B. acknowledges support by a Center of Excellence of the Israel Science Foundation (grant No.2752/19).
The Australia Telescope is funded by the Commonwealth of Australia for operation as a National Facility managed by CSIRO.
The National Radio Astronomy Observatory is a facility of the National Science Foundation operated under cooperative agreement by Associated Universities, Inc.

\section*{Data availability}

The data underlying this article are in the Australia Telescope Online Archive (ATOA) at \url{https://atoa.atnf.csiro.au/query.jsp} with project codes of C3329 and CX414.

\bibliographystyle{mnras}
\bibliography{atca.bbl}

\onecolumn

\setcounter{table}{0}

\begin{longtable}{cccccccc}
\caption{The ATCA observation details.} \label{observation} \\
\hline
\hline
\multirow{2}{*}{Name} & R.A. & Dec. & \multirow{2}{*}{$z$} & Obs. date & Exp. time & RMS$_{\rm{5.5GHz}}$ & RMS$_{\rm{9.0GHz}}$ \\
& (HH:MM:SS) & (DD:MM:SS) & & (YYYY-MM-DD) & (minute) & (mJy beam$^{-1}$) & (mJy beam$^{-1}$) \\
(1) & (2) & (3) & (4) & (5) & (6) & (7) & (8) \\
\hline
\endfirsthead
\caption{continued.} \\
\hline
\hline
\multirow{2}{*}{Name} & R.A. & Dec. & \multirow{2}{*}{$z$} & Obs. date & Exp. time & RMS$_{\rm{5.5GHz}}$ & RMS$_{\rm{9.0GHz}}$ \\
& (HH:MM:SS) & (DD:MM:SS) & & (YYYY-MM-DD) & (minute) & (mJy~beam$^{-1}$) & (mJy~beam$^{-1}$) \\
(1) & (2) & (3) & (4) & (5) & (6) & (7) & (8) \\
\hline
\endhead
\hline
\endfoot
\multicolumn{8}{c}{ATCA 6km array} \\
\hline
J0013$-$3238 & 00:13:02.25 & $-$32:38:29.9 & 0.259 & 2020-10-28 & 80 & 0.03 & 0.03 \\
J0016$-$3126 & 00:16:41.86 & $-$31:26:56.9 & 0.360 & 2020-10-27 & 80 & 0.03 & 0.03 \\
J0032$-$3242 & 00:32:10.95 & $-$32:42:05.1 & 0.354 & 2020-10-27 & 80 & 0.04 & 0.03 \\
J0040$-$3713 & 00:40:39.17 & $-$37:13:16.8 & 0.036 & 2020-10-28 & 80 & 0.03 & 0.03 \\
J0117$-$3826 & 01:17:30.64 & $-$38:26:29.9 & 0.225 & 2020-10-28 & 80 & 0.04 & 0.03 \\
J0119$-$2821 & 01:19:35.69 & $-$28:21:31.5 & 0.349 & 2020-10-28 & 80 & 0.04 & 0.03 \\
J0131$-$3956 & 01:31:17.09 & $-$39:56:30.7 & 0.499 & 2020-10-28 & 80 & 0.03 & 0.03 \\
J0135$-$3539 & 01:35:46.35 & $-$35:39:15.3 & 0.137 & 2020-09-24 & 20 & 0.08 & 0.06 \\
J0142$-$3525 & 01:42:22.28 & $-$35:25:41.4 & 0.094 & 2020-10-28 & 80 & 0.03 & 0.03 \\
J0217$-$3004 & 02:17:38.10 & $-$30:04:47.7 & 0.080 & 2020-10-28 & 80 & 0.03 & 0.03 \\
J0315$-$2901 & 03:15:58.92 & $-$29:01:26.4 & 0.307 & 2020-04-10 & 60 & 0.04 & 0.05 \\
J0420$-$3551 & 04:20:20.40 & $-$35:51:00.3 & 0.196 & 2020-04-10 & 60 & 0.04 & 0.04 \\
J0436$-$3010 & 04:36:59.14 & $-$30:10:41.9 & 0.100 & 2020-04-28 & 80 & 0.03 & 0.02 \\
J0523$-$3305 & 05:23:49.52 & $-$33:05:50.5 & 0.203 & 2020-05-08 & 80 & 0.03 & 0.02 \\
J0537$-$3817 & 05:37:24.48 & $-$38:17:58.7 & 0.332 & 2020-04-28 & 80 & 0.03 & 0.03 \\
J0541$-$3738 & 05:41:58.03 & $-$37:38:37.1 & 0.224 & 2020-04-28 & 80 & 0.03 & 0.02 \\
J0554$-$3453 & 05:54:26.42 & $-$34:53:40.6 & 0.082 & 2020-05-08 & 80 & 0.03 & 0.02 \\
J0650$-$3805 & 06:50:17.46 & $-$38:05:13.5 & 0.030 & 2020-04-10 & 60 & 0.03 & 0.03 \\
J1032$-$2707 & 10:32:57.04 & $-$27:07:30.3 & 0.071 & 2020-04-28 & 20 & 0.06 & 0.05 \\
J1042$-$3800 & 10:42:03.00 & $-$38:00:57.8 & 0.117 & 2020-04-28 & 60 & 0.03 & 0.03 \\
J1048$-$3902 & 10:48:33.84 & $-$39:02:37.9 & 0.045 & 2020-04-10 & 60 & 0.04 & 0.04 \\
J1143$-$3615 & 11:43:53.82 & $-$36:15:18.4 & 0.102 & 2020-04-28 & 60 & 0.04 & 0.03 \\
J1238$-$3440 & 12:38:17.20 & $-$34:40:28.2 & 0.076 & 2020-04-28 & 80 & 0.03 & 0.02 \\
J1321$-$3104 & 13:21:58.19 & $-$31:04:25.7 & 0.045 & 2020-05-08 & 80 & 0.03 & 0.02 \\
J1325$-$3824 & 13:25:19.38 & $-$38:24:52.7 & 0.066 & 2020-04-28 & 80 & 0.03 & 0.02 \\
J1329$-$2925 & 13:29:04.60 & $-$29:25:56.1 & 0.302 & 2020-05-08 & 80 & 0.03 & 0.02 \\
J1340$-$3521 & 13:40:54.24 & $-$35:21:11.0 & 0.278 & 2020-05-08 & 80 & 0.03 & 0.03 \\
J1407$-$3120 & 14:07:22.00 & $-$31:20:57.8 & 0.075 & 2020-09-24 & 80 & 0.04 & 0.03 \\
J1410$-$3644 & 14:10:48.73 & $-$36:44:11.9 & 0.063 & 2020-05-08 & 20 & 0.07 & 0.06 \\
J1447$-$3802 & 14:47:18.72 & $-$38:02:21.8 & 0.064 & 2020-09-24 & 80 & 0.03 & 0.02 \\
J1512$-$3333 & 15:12:22.47 & $-$33:33:34.2 & 0.024 & 2020-05-08 & 20 & 0.07 & 0.05 \\
J2039$-$3018 & 20:39:27.19 & $-$30:18:52.2 & 0.079 & 2020-09-24 & 20 & 0.06 & 0.05 \\
J2045$-$3010 & 20:45:13.10 & $-$30:10:26.9 & 0.111 & 2020-09-25 & 70 & 0.04 & 0.03 \\
J2059$-$3147 & 20:59:20.74 & $-$31:47:34.8 & 0.073 & 2020-09-24 & 20 & 0.06 & 0.05 \\
J2133$-$3558 & 21:33:33.36 & $-$35:58:48.4 & 0.350 & 2020-10-27 & 80 & 0.03 & 0.03 \\
J2157$-$3123 & 21:57:23.17 & $-$31:23:06.5 & 0.085 & 2020-10-27 & 80 & 0.03 & 0.03 \\
J2216$-$3947 & 22:16:11.23 & $-$39:47:33.7 & 0.247 & 2020-10-27 & 80 & 0.03 & 0.03 \\
J2218$-$3957 & 22:18:00.19 & $-$39:57:22.7 & 0.244 & 2020-10-27 & 80 & 0.04 & 0.04 \\
J2242$-$3845 & 22:42:37.66 & $-$38:45:16.3 & 0.220 & 2020-10-27 & 80 & 0.03 & 0.03 \\
\hline
\multicolumn{8}{c}{ATCA 750m array} \\
\hline
J0037$-$5420 & 00:37:40.00 & $-$54:20:29.9 & 0.101 & 2018-09-17,26,27 & 58 & 0.03 & 0.03 \\
J0039$-$5117 & 00:39:15.85 & $-$51:17:01.5 & 0.029 & 2018-09-17,26,27 & 60 & 0.04 & 0.03 \\
J0103$-$4804 & 01:03:13.19 & $-$48:04:45.1 & 0.424 & 2018-09-17,26,27 & 60 & 0.04 & 0.03 \\
J0114$-$5136 & 01:14:42.46 & $-$51:36:13.8 & 0.205 & 2018-09-17,26,27 & 58 & 0.03 & 0.03 \\
J0152$-$5238 & 01:52:51.57 & $-$52:38:24.3 & 0.077 & 2018-09-17,26,27 & 58 & 0.03 & 0.03 \\
J0222$-$5332 & 02:22:19.44 & $-$53:32:39.1 & 0.095 & 2018-09-17,26,27 & 58 & 0.04 & 0.03 \\
J0228$-$4057 & 02:28:15.23 & $-$40:57:14.7 & 0.493 & 2018-09-17,26,27 & 60 & 0.03 & 0.03 \\
J0239$-$5345 & 02:39:27.62 & $-$53:45:07.9 & 0.312 & 2018-09-17,26,27 & 58 & 0.03 & 0.03 \\
J0252$-$5922 & 02:52:28.38 & $-$59:22:43.0 & 0.209 & 2018-09-17,26,27 & 77 & 0.06 & 0.03 \\
J0307$-$7250 & 03:07:35.32 & $-$72:50:02.5 & 0.028 & 2018-09-17,26,27 & 80 & 0.04 & 0.03 \\
J0440$-$4110 & 04:40:40.35 & $-$41:10:43.7 & 0.033 & 2018-09-17,26,27 & 80 & 0.03 & 0.03 \\
J0507$-$4652 & 05:07:43.97 & $-$46:52:30.6 & 0.121 & 2018-09-17,26,27 & 77 & 0.03 & 0.03 \\
J0522$-$4222 & 05:22:10.06 & $-$42:22:54.7 & 0.261 & 2018-09-17,26,27 & 80 & 0.03 & 0.03 \\
J0609$-$5606 & 06:09:17.48 & $-$56:06:58.4 & 0.032 & 2018-09-17,26,27 & 40 & 0.04 & 0.04 \\
J0615$-$5826 & 06:15:49.59 & $-$58:26:05.1 & 0.055 & 2018-09-17,26,27 & 40 & 0.04 & 0.04 \\
J0639$-$5125 & 06:39:40.77 & $-$51:25:14.5 & 0.108 & 2018-09-17,26,27 & 40 & 0.04 & 0.03 \\
J0641$-$4317 & 06:41:57.71 & $-$43:17:41.8 & 0.061 & 2018-09-17,26,27 & 31 & 0.04 & 0.03 \\
J0656$-$6315 & 06:56:54.25 & $-$63:15:35.5 & 0.036 & 2018-09-17,26,27 & 29 & 0.03 & 0.03 \\
J0708$-$4933 & 07:08:41.50 & $-$49:33:06.4 & 0.041 & 2018-09-17,26,27 & 32 & 0.04 & 0.04 \\
J0710$-$5357 & 07:10:32.99 & $-$53:57:52.3 & 0.048 & 2018-09-17,26,27 & 29 & 0.03 & 0.04 \\
J0842$-$7042 & 08:42:34.99 & $-$70:42:48.5 & 0.110 & 2018-09-17,26,27 & 39 & 0.04 & 0.03 \\
J1011$-$4528 & 10:11:34.82 & $-$45:28:29.8 & 0.057 & 2018-09-17,19,26,27 & 43 & 0.03 & 0.03 \\
J1042$-$4000 & 10:42:08.92 & $-$40:00:31.6 & 0.386 & 2018-09-17,19,26,27 & 35 & 0.04 & 0.03 \\
J1057$-$4039 & 10:57:27.87 & $-$40:39:40.6 & 0.398 & 2018-09-19,26,27 & 33 & 0.06 & 0.07 \\
J1257$-$4435 & 12:57:06.43 & $-$44:35:20.0 & 0.097 & 2018-09-19,26,27 & 45 & 0.04 & 0.03 \\
J1354$-$4214 & 13:54:39.48 & $-$42:14:56.5 & 0.076 & 2018-09-19,26,27 & 37 & 0.04 & 0.03 \\
J1434$-$4254 & 14:34:38.43 & $-$42:54:05.3 & 0.115 & 2018-09-19,26,27 & 33 & 0.04 & 0.04 \\
J1500$-$7248 & 15:00:12.81 & $-$72:48:40.3 & 0.141 & 2018-09-17,26,27 & 79 & 0.04 & 0.04 \\
J1515$-$7820 & 15:15:15.20 & $-$78:20:12.0 & 0.259 & 2018-09-17,26,27 & 77 & 0.04 & 0.03 \\
J1729$-$6854 & 17:29:16.25 & $-$68:54:41.5 & 0.048 & 2018-09-17,26,27 & 80 & 0.04 & 0.03 \\
J1834$-$5724 & 18:34:50.81 & $-$57:24:22.1 & 0.055 & 2018-09-17,26,27 & 80 & 0.04 & 0.03 \\
J1849$-$5513 & 18:49:42.12 & $-$55:13:09.5 & 0.050 & 2018-09-17,26,27 & 70 & 0.03 & 0.03 \\
J1938$-$4326 & 19:38:19.58 & $-$43:26:46.2 & 0.079 & 2018-09-17,26,27 & 68 & 0.03 & 0.03 \\
J1957$-$4141 & 19:57:05.25 & $-$41:41:16.9 & 0.374 & 2018-09-17,26,27 & 78 & 0.04 & 0.03 \\
J2005$-$4134 & 20:05:53.00 & $-$41:34:42.1 & 0.080 & 2018-09-17,26,27 & 78 & 0.03 & 0.03 \\
J2025$-$4822 & 20:25:57.37 & $-$48:22:26.4 & 0.067 & 2018-09-17,26,27 & 60 & 0.07 & 0.05 \\
J2059$-$5136 & 20:59:33.04 & $-$51:36:00.5 & 0.314 & 2018-09-17,26,27 & 59 & 0.04 & 0.03 \\
J2112$-$4128 & 21:12:24.60 & $-$41:28:53.8 & 0.349 & 2018-09-17,26,27 & 60 & 0.07 & 0.03 \\
J2135$-$6230 & 21:35:29.50 & $-$62:30:07.2 & 0.061 & 2018-09-17,26,27 & 60 & 0.03 & 0.03 \\
J2145$-$6922 & 21:45:05.75 & $-$69:22:30.9 & 0.089 & 2018-09-17,26,27 & 60 & 0.06 & 0.04 \\
J2216$-$4451 & 22:16:53.21 & $-$44:51:57.0 & 0.135 & 2018-09-17,26,27 & 59 & 0.04 & 0.03 \\
J2245$-$4652 & 22:45:20.30 & $-$46:52:11.4 & 0.200 & 2018-09-17,26,27 & 58 & 0.04 & 0.03 \\
J2253$-$5118 & 22:53:21.60 & $-$51:18:02.2 & 0.115 & 2018-09-17,26,27 & 58 & 0.03 & 0.03 \\
J2340$-$5941 & 23:40:42.00 & $-$59:41:32.3 & 0.379 & 2018-09-17,26,27 & 59 & 0.03 & 0.03 \\
J2341$-$5923 & 23:41:20.64 & $-$59:23:58.6 & 0.203 & 2018-09-17,26,27 & 60 & 0.03 & 0.03 \\
J2346$-$4909 & 23:46:14.64 & $-$49:09:51.8 & 0.203 & 2018-09-17,26,27 & 58 & 0.03 & 0.03 \\
\hline
\end{longtable}
\begin{flushleft}
\textbf{Notes.} Columns: (1) name, (2) right ascension, (3) declination, (4) redshift, (5) observing date, (6) exposure time, (7) RMS at 5.5~GHz, (8) RMS at 9.0~GHz.
\end{flushleft}

\setcounter{table}{1}

\begin{table}
\centering
\caption{The sizes of the ATCA-detected sources.}
\begin{footnotesize}
\begin{tabular}{ccccccccc}
\hline
\hline
\multirow{3}{*}{Name} & \multirow{3}{*}{Array} & \multirow{2}{*}{Scale} & \multicolumn{3}{c}{5.5~GHz} & \multicolumn{3}{c}{9.0~GHz} \\
& & & Core Maj & Core Min & Core PA & Core Maj & Core Min & Core PA \\
& & (kpc~arcsec$^{-1}$) & (kpc) & (kpc) & (degree) & (kpc) & (kpc) & (degree) \\
(1) & (2) & (3) & (4) & (5) & (6) & (7) & (8) & (9) \\
\hline
J0016$-$3126 & 6km & 9.32 & $<$ 10.2 & $<$ 4.94 & - & $<$ 6.52 & $<$ 3.12 & - \\
J0040$-$3713 & 6km & 0.77 & $<$ 0.73 & $<$ 0.43 & - & $<$ 0.43 & $<$ 0.28 & - \\
J0135$-$3539 & 6km & 3.14 & 60.87 & 2.89 & 113.0 & $<$ 28.68 & $<$ 1.13 & - \\
J0142$-$3525 & 6km & 2.09 & $<$ 2.07 & $<$ 1.17 & - & $<$ 1.22 & $<$ 0.75 & - \\
J0152$-$5238 & 750m & 1.68 & $<$ 24.47 & $<$ 0.52 & - & - & - & - \\
J0217$-$3004 & 6km & 1.77 & $<$ 2.01 & $<$ 1.0 & - & $<$ 1.18 & $<$ 0.65 & - \\
J0222$-$5332 & 750m & 2.12 & $<$ 24.43 & $<$ 0.65 & - & 13.97 & 0.38 & 27.5 \\
J0228$-$4057 & 750m & 13.52 & 124.35 & 3.51 & 21.6 & - & - & - \\
J0307$-$7250 & 750m & 0.59 & - & - & - & 5.7 & 0.12 & 36.5 \\
J0436$-$3010 & 6km & 2.23 & $<$ 5.74 & $<$ 0.99 & - & $<$ 3.63 & $<$ 0.62 & - \\
J0440$-$4110 & 750m & 0.7 & $<$ 8.09 & $<$ 0.2 & - & 5.06 & 0.2 & 11.9 \\
J0507$-$4652 & 750m & 2.73 & 11.85 & 1.23 & 12.2 & 18.52 & 0.87 & 11.4 \\
J0537$-$3817 & 6km & 8.45 & $<$ 21.8 & $<$ 2.92 & - & $<$ 13.44 & $<$ 1.86 & - \\
J0541$-$3738 & 6km & 5.39 & $<$ 12.5 & $<$ 2.21 & - & - & - & - \\
J0554$-$3453 & 6km & 1.8 & $<$ 5.92 & $<$ 0.7 & - & $<$ 3.73 & $<$ 0.44 & - \\
J0609$-$5606 & 750m & 0.68 & 11.38 & 0.28 & 6.2 & - & - & - \\
J0842$-$7042 & 750m & 2.46 & 37.53 & 0.37 & 170.5 & - & - & - \\
J1032$-$2707 & 6km & 1.56 & 32.32 & 0.5 & 179.9 & 29.35 & 0.16 & 180.0 \\
J1042$-$4000 & 750m & 10.1 & $<$ 25.04 & $<$ 4.49 & - & $<$ 15.2 & $<$ 2.78 & - \\
J1048$-$3902 & 6km & 0.96 & $<$ 1.15 & $<$ 0.44 & - & $<$ 0.73 & $<$ 0.28 & - \\
J1057$-$4039 & 750m & 10.47 & 17.28 & 3.56 & 137.1 & 11.42 & 2.09 & 138.3 \\
J1257$-$4435 & 750m & 2.15 & 19.95 & 0.58 & 114.7 & 16.68 & 0.54 & 114.6 \\
J1321$-$3104 & 6km & 0.97 & $<$ 2.05 & $<$ 0.4 & - & $<$ 1.29 & $<$ 0.25 & - \\
J1325$-$3824 & 6km & 1.43 & $<$ 2.62 & $<$ 0.55 & - & $<$ 1.66 & $<$ 0.35 & - \\
J1329$-$2925 & 6km & 7.59 & $<$ 17.03 & $<$ 3.04 & - & $<$ 10.77 & $<$ 1.93 & - \\
J1407$-$3120 & 6km & 1.64 & $<$ 2.94 & $<$ 0.84 & - & $<$ 1.86 & $<$ 0.52 & - \\
J1410$-$3644 & 6km & 1.36 & $<$ 1.56 & $<$ 0.7 & - & $<$ 0.99 & $<$ 0.44 & - \\
J1434$-$4254 & 750m & 2.58 & 1.45 & 0.26 & 61.7 & - & - & - \\
J1447$-$3802 & 6km & 1.39 & $<$ 2.36 & $<$ 0.62 & - & $<$ 1.5 & $<$ 0.4 & - \\
J1500$-$7248 & 750m & 3.24 & 5.57 & 1.3 & 117.5 & 719.45 & 573.94 & 73.0 \\
J1512$-$3333 & 6km & 0.5 & 2.56 & 0.18 & 130.7 & 3.44 & 0.16 & 130.6 \\
J1515$-$7820 & 750m & 6.35 & 11.94 & 1.21 & 115.8 & 13.08 & 0.89 & 115.0 \\
J1729$-$6854 & 750m & 1.04 & $<$ 0.84 & $<$ 0.45 & - & - & - & - \\
J1938$-$4326 & 750m & 1.74 & $<$ 1.97 & $<$ 0.81 & - & - & - & - \\
J2039$-$3018 & 6km & 1.74 & $<$ 4.21 & $<$ 0.9 & - & $<$ 2.67 & $<$ 0.57 & - \\
J2045$-$3010 & 6km & 2.5 & 23.41 & 1.25 & 167.9 & $<$ 14.56 & $<$ 0.51 & - \\
J2059$-$3147 & 6km & 1.61 & 3.6 & 1.4 & 127.0 & 7.85 & 0.76 & 149.2 \\
J2133$-$3558 & 6km & 8.99 & $<$ 9.67 & $<$ 4.41 & - & - & - & - \\
J2157$-$3123 & 6km & 1.88 & $<$ 2.27 & $<$ 0.92 & - & 2.46 & 0.73 & 172.6 \\
J2216$-$3947 & 6km & 6.03 & $<$ 5.33 & $<$ 3.22 & - & 6.99 & 0.72 & 24.0 \\
J2245$-$4652 & 750m & 4.75 & 38.87 & 1.33 & 42.8 & 45.1 & 0.81 & 42.9 \\
J2340$-$5941 & 750m & 9.88 & $<$ 24.94 & $<$ 3.31 & - & 106.2 & 1.78 & 41.0 \\
\hline
\end{tabular}
\end{footnotesize}
\label{size}
\flushleft{\textbf{Notes.} Columns: (1) name, (2) array, (3) scale, (4) core major axis at 5.5~GHz, (5) core minor axis at 5.5~GHz, (6) core position angle at 5.5~GHz, (7) core major axis at 9.0~GHz, (8) core minor axis at 9.0~GHz, (9) core position angle at 9.0~GHz. Some objects have very elongated beam because they were only observed for one scan or cut due to RFI.}
\end{table}

\setcounter{table}{2}

\begin{table}
\centering
\caption{The radio flux densities in different bands of the ATCA-detected sources.}
\begin{footnotesize}
\begin{tabular}{cccccc}
\hline
\hline
\multirow{2}{*}{Name} & $S_{\rm 9.0GHz}$ & $S_{\rm 5.5GHz}$ & $S_{\rm 3.0GHz}$ & $S_{\rm 1.4GHz}$ & $S_{\rm 843MHz}$ \\
& (mJy) & (mJy) & (mJy) & (mJy) & (mJy) \\
(1) & (2) & (3) & (4) & (5) & (6) \\
\hline
J0016$-$3126 & 0.13 $\pm$ 0.03 & 0.23 $\pm$ 0.03 & < 0.36 & < 1.35 & < 10 \\
J0040$-$3713 & 0.43 $\pm$ 0.03 & 0.7 $\pm$ 0.03 & 1.02 $\pm$ 0.16 & < 1.35 & < 10 \\
J0135$-$3539 & 0.3 $\pm$ 0.06 & 0.9 $\pm$ 0.08 & 1.6 $\pm$ 0.31 & 2.2 $\pm$ 0.5 & < 10 \\
J0142$-$3525 & 0.14 $\pm$ 0.03 & 0.18 $\pm$ 0.03 & < 0.36 & < 1.35 & < 10 \\
J0152$-$5238 & < 0.09 & 0.4 $\pm$ 0.03 & - & - & < 6 \\
J0217$-$3004 & 0.16 $\pm$ 0.03 & 0.29 $\pm$ 0.03 & 0.39 $\pm$ 0.04 & < 1.35 & < 10 \\
J0222$-$5332 & 0.4 $\pm$ 0.03 & 0.7 $\pm$ 0.04 & - & - & < 6 \\
J0228$-$4057 & < 0.09 & 0.6 $\pm$ 0.03 & - & - & < 10 \\
J0307$-$7250 & 0.4 $\pm$ 0.03 & < 0.12 & - & - & < 6 \\
J0436$-$3010 & 1.1 $\pm$ 0.02 & 1.03 $\pm$ 0.03 & 1.15 $\pm$ 0.19 & < 1.35 & < 10 \\
J0440$-$4110 & 0.4 $\pm$ 0.03 & 0.7 $\pm$ 0.03 & - & - & < 10 \\
J0507$-$4652 & 0.6 $\pm$ 0.03 & 1.1 $\pm$ 0.03 & - & - & < 10 \\
J0537$-$3817 & 0.14 $\pm$ 0.03 & 0.15 $\pm$ 0.03 & < 0.36 & < 1.35 & < 10 \\
J0541$-$3738 & < 0.06 & 0.26 $\pm$ 0.03 & 0.71 $\pm$ 0.1 & < 1.35 & < 10 \\
J0554$-$3453 & 0.11 $\pm$ 0.02 & 0.16 $\pm$ 0.03 & < 0.36 & < 1.35 & < 10 \\
J0609$-$5606 & < 0.12 & 0.4 $\pm$ 0.04 & - & - & 6.8 $\pm$ 1.0 \\
J0842$-$7042 & < 0.09 & 0.5 $\pm$ 0.04 & - & - & < 6 \\
J1032$-$2707 & 0.72 $\pm$ 0.05 & 0.93 $\pm$ 0.06 & 2.54 $\pm$ 0.53 & 5.3 $\pm$ 0.6 & - \\
J1042$-$4000 & 1.3 $\pm$ 0.03 & 2.3 $\pm$ 0.04 & 3.77 $\pm$ 0.25 & 5.7 $\pm$ 0.5 & < 10 \\
J1048$-$3902 & 0.29 $\pm$ 0.04 & 0.42 $\pm$ 0.04 & 0.72 $\pm$ 0.21 & < 1.35 & < 10 \\
J1057$-$4039 & 151.7 $\pm$ 0.07 & 167.7 $\pm$ 0.06 & - & - & 437.5 $\pm$ 13.2 \\
J1257$-$4435 & 1.1 $\pm$ 0.03 & 1.8 $\pm$ 0.04 & - & - & < 10 \\
J1321$-$3104 & 0.25 $\pm$ 0.02 & 0.36 $\pm$ 0.03 & < 0.36 & < 1.35 & < 10 \\
J1325$-$3824 & 1.14 $\pm$ 0.02 & 2.01 $\pm$ 0.03 & 3.89 $\pm$ 0.69 & 6.3 $\pm$ 0.6 & < 10 \\
J1329$-$2925 & 0.29 $\pm$ 0.02 & 0.44 $\pm$ 0.03 & 0.58 $\pm$ 0.18 & < 1.35 & - \\
J1407$-$3120 & 0.26 $\pm$ 0.03 & 0.57 $\pm$ 0.04 & < 0.36 & < 1.35 & < 10 \\
J1410$-$3644 & 0.29 $\pm$ 0.06 & 0.55 $\pm$ 0.07 & 1.3 $\pm$ 0.21 & 2.9 $\pm$ 0.5 & < 10 \\
J1434$-$4254 & < 0.12 & 2.3 $\pm$ 0.04 & - & - & 10.5 $\pm$ 1.2 \\
J1447$-$3802 & 0.24 $\pm$ 0.02 & 0.28 $\pm$ 0.03 & < 0.36 & < 1.35 & < 10 \\
J1500$-$7248 & 35.4 $\pm$ 0.04 & 31.4 $\pm$ 0.04 & - & - & 20.1 $\pm$ 1.5 \\
J1512$-$3333 & 2.03 $\pm$ 0.05 & 2.89 $\pm$ 0.07 & 4.91 $\pm$ 0.48 & 9.0 $\pm$ 0.6 & 12.0 $\pm$ 1.6 \\
J1515$-$7820 & 2.1 $\pm$ 0.03 & 3.9 $\pm$ 0.04 & - & - & 33.0 $\pm$ 1.8 \\
J1729$-$6854 & < 0.09 & 0.5 $\pm$ 0.04 & - & - & < 6 \\
J1938$-$4326 & < 0.09 & 0.6 $\pm$ 0.03 & - & - & < 10 \\
J2039$-$3018 & 1.01 $\pm$ 0.05 & 1.99 $\pm$ 0.06 & 3.59 $\pm$ 0.32 & 5.8 $\pm$ 0.5 & < 10 \\
J2045$-$3010 & 0.18 $\pm$ 0.03 & 0.22 $\pm$ 0.04 & < 0.36 & < 1.35 & < 10 \\
J2059$-$3147 & 0.38 $\pm$ 0.05 & 1.09 $\pm$ 0.06 & 5.02 $\pm$ 0.65 & 9.6 $\pm$ 0.6 & 12.3 $\pm$ 1.3 \\
J2133$-$3558 & < 0.09 & 0.13 $\pm$ 0.03 & < 0.36 & < 1.35 & < 10 \\
J2157$-$3123 & 0.21 $\pm$ 0.03 & 0.28 $\pm$ 0.03 & < 0.36 & < 1.35 & < 10 \\
J2216$-$3947 & 0.18 $\pm$ 0.03 & 0.41 $\pm$ 0.03 & 0.81 $\pm$ 0.21 & < 1.35 & < 10 \\
J2245$-$4652 & 1.6 $\pm$ 0.03 & 2.6 $\pm$ 0.04 & - & - & < 10 \\
J2340$-$5941 & 0.5 $\pm$ 0.03 & 0.8 $\pm$ 0.03 & - & - & < 6 \\
\hline
Median & 0.40 $\pm$ 0.03 & 0.60 $\pm$ 0.03 & 1.30 $\pm$ 0.21 & 5.75 $\pm$ 0.55 & 12.3 $\pm$ 1.5 \\
\hline
\end{tabular}
\end{footnotesize}
\label{flux}
\flushleft{\textbf{Notes.} Columns: (1) name, (2) flux density at 9.0~GHz from ATCA, (3) flux density at 5.5~GHz from ATCA, (4) flux density at 3.0~GHz from VLASS, (5) flux density at 1.4~GHz from NVSS, (6) flux density at 843~MHz from SUMSS. The last line shows the median values of the detected flux densities at different frequencies.}
\end{table}

\setcounter{table}{3}

\begin{table}
\centering
\caption{The spectral slopes at 0.8--9.0~GHz for ATCA-detected objects}
\begin{footnotesize}
\begin{tabular}{ccccccc}
\hline
\hline
Name & $\alpha_{5.5-9.0}$ & $\alpha_{3.0-5.5}$ & $\alpha_{1.4-3.0}$ & $\alpha_{0.8-1.4}$ & $\alpha_{0.8-5.5}$ & Radio emission \\
(1) & (2) & (3) & (4) & (5) & (6) & (7) \\
\hline
J0016$-$3126 & $-$1.16 $\pm$ 0.54 & $>$ $-$0.74 & - & - & - & - \\
J0040$-$3713 & $-$0.99 $\pm$ 0.17 & $-$0.62 $\pm$ 0.27 & $>$ $-$0.37 & - & - & Steep \\
J0135$-$3539 & $-$2.23 $\pm$ 0.44 & $-$0.95 $\pm$ 0.35 & $-$0.42 $\pm$ 0.39 & - & - & Relic \\
J0142$-$3525 & $-$0.51 $\pm$ 0.55 & $>$ $-$1.14 & - & - & - & - \\
J0152$-$5238 & $<$ $-$3.03 & - & - & - & - & - \\
J0217$-$3004 & $-$1.21 $\pm$ 0.43 & $-$0.49 $\pm$ 0.24 & $>$ $-$1.63 & - & - & Steep \\
J0222$-$5332 & $-$1.14 $\pm$ 0.19 & - & - & - & - & - \\
J0228$-$4057 & $<$ $-$3.85 & - & - & - & - & - \\
J0307$-$7250 & $>$ 2.44 & - & - & - & - & - \\
J0436$-$3010 & 0.13 $\pm$ 0.07 & $-$0.18 $\pm$ 0.28 & $>$ $-$0.21 & - & - & Flat \\
J0440$-$4110 & $-$1.14 $\pm$ 0.18 & - & - & - & - & - \\
J0507$-$4652 & $-$1.23 $\pm$ 0.12 & - & - & - & - & - \\
J0537$-$3817 & $-$0.14 $\pm$ 0.6 & $>$ $-$1.44 & - & - & - & - \\
J0541$-$3738 & $<$ $-$2.98 & $-$1.66 $\pm$ 0.3 & $>$ $-$0.84 & - & - & - \\
J0554$-$3453 & $-$0.76 $\pm$ 0.53 & $>$ $-$1.34 & - & - & - & - \\
J0609$-$5606 & $<$ $-$2.44 & - & - & - & $-$1.51 $\pm$ 0.09 & - \\
J0842$-$7042 & $<$ $-$3.48 & - & - & - & - & - \\
J1032$-$2707 & $-$0.52 $\pm$ 0.19 & $-$1.66 $\pm$ 0.36 & $-$0.97 $\pm$ 0.31 & - & - & HFE \\
J1042$-$4000 & $-$1.16 $\pm$ 0.06 & $-$0.82 $\pm$ 0.11 & $-$0.54 $\pm$ 0.14 & - & - & Steep \\
J1048$-$3902 & $-$0.75 $\pm$ 0.34 & $-$0.89 $\pm$ 0.51 & $>$ $-$0.82 & - & - & SF \\
J1057$-$4039 & $-$0.2 $\pm$ 0.01 & - & - & - & $-$0.51 $\pm$ 0.02 & Flat \\
J1257$-$4435 & $-$1.0 $\pm$ 0.07 & - & - & - & - & - \\
J1321$-$3104 & $-$0.74 $\pm$ 0.23 & $>$ 0.01 & - & - & - & - \\
J1325$-$3824 & $-$1.15 $\pm$ 0.05 & $-$1.09 $\pm$ 0.29 & $-$0.63 $\pm$ 0.26 & - & - & Steep \\
J1329$-$2925 & $-$0.85 $\pm$ 0.2 & $-$0.46 $\pm$ 0.52 & $>$ $-$1.11 & - & - & Steep \\
J1407$-$3120 & $-$1.59 $\pm$ 0.27 & $>$ 0.76 & - & - & - & - \\
J1410$-$3644 & $-$1.3 $\pm$ 0.49 & $-$1.42 $\pm$ 0.34 & $-$1.05 $\pm$ 0.31 & - & - & Steep \\
J1434$-$4254 & $<$ $-$6.0 & - & - & - & $-$0.81 $\pm$ 0.06 & - \\
J1447$-$3802 & $-$0.31 $\pm$ 0.28 & $>$ $-$0.41 & - & - & - & - \\
J1500$-$7248 & 0.24 $\pm$ 0.01 & - & - & - & 0.24 $\pm$ 0.04 & Flat \\
J1512$-$3333 & $-$0.72 $\pm$ 0.07 & $-$0.87 $\pm$ 0.17 & $-$0.8 $\pm$ 0.16 & $-$0.57 $\pm$ 0.29 & - & SF \\
J1515$-$7820 & $-$1.26 $\pm$ 0.04 & - & - & - & $-$1.14 $\pm$ 0.03 & Steep \\
J1729$-$6854 & $<$ $-$3.48 & - & - & - & - & - \\
J1938$-$4326 & $<$ $-$3.85 & - & - & - & - & - \\
J2039$-$3018 & $-$1.38 $\pm$ 0.12 & $-$0.97 $\pm$ 0.16 & $-$0.63 $\pm$ 0.16 & - & - & Steep \\
J2045$-$3010 & $-$0.41 $\pm$ 0.5 & $>$ $-$0.81 & - & - & - & - \\
J2059$-$3147 & $-$2.14 $\pm$ 0.29 & $-$2.52 $\pm$ 0.23 & $-$0.85 $\pm$ 0.19 & $-$0.49 $\pm$ 0.24 & - & Relic \\
J2133$-$3558 & $<$ $-$0.75 & $>$ $-$1.68 & - & - & - & - \\
J2157$-$3123 & $-$0.58 $\pm$ 0.36 & $>$ $-$0.41 & - & - & - & - \\
J2216$-$3947 & $-$1.67 $\pm$ 0.37 & $-$1.12 $\pm$ 0.44 & $>$ $-$0.67 & - & - & Steep \\
J2245$-$4652 & $-$0.99 $\pm$ 0.05 & - & - & - & - & - \\
J2340$-$5941 & $-$0.95 $\pm$ 0.14 & - & - & - & - & - \\
\hline
Median & $-$0.99 $\pm$ 0.20 & $-$0.95 $\pm$ 0.29 & $-$0.72 $\pm$ 0.23 & $-$0.53 $\pm$ 0.27 & $-$0.81 $\pm$ 0.04 & - \\
\hline
\end{tabular}
\end{footnotesize}
\label{slope}
\flushleft{\textbf{Notes.} Columns: (1) name, (2) spectral slope at 5.5--9.0~GHz, (3) spectral slope at 3.0--5.5~GHz, (4) spectral slope at 1.4--3.0~GHz, (5) spectral slope at 0.8--1.4~GHz, (6) spectral slope at 0.8--5.5~GHz, (7) the origin of the radio emission of the 17 NLS1s analyzed in Section 4.1.
Flat: a flat or inverted slope may indicate a jet or a corona, Steep: a slope steepens with increasing frequencies may demonstrate an AGN outflow or a jet, HFE: high frequency excess may be present, Relic: relic emission from past AGN activity may be present, SF: radio emission may be associated with star formation.
The last line shows the median values of the detected spectral indices at different frequencies.}
\end{table}

\setcounter{table}{4}

\begin{table}
\centering
\caption{The multi-wavelength properties of the ATCA-detected sources.}
\begin{footnotesize}
\begin{tabular}{cccccccc}
\hline
\hline
\multirow{2}{*}{Name} & \multirow{2}{*}{$R_{\rm L}$} & $M_{\rm B}$ & \multirow{2}{*}{$\log \rm{FWHM}(\rm{H}\beta)$} & \multirow{2}{*}{$R_{\rm{Fe~II}/\rm{H}\beta}$} & $\log L_{5100\text{\AA}}$ & \multirow{2}{*}{$\log M_{\rm{BH}}/M_{\odot}$} & \multirow{2}{*}{$\log L_{\rm{bol}}/L_{\rm{Edd}}$} \\
& & (mag) & & & (erg~s$^{-1}$) & & \\
(1) & (2) & (3) & (4) & (5) & (6) & (7) & (8) \\
\hline
J0016$-$3126 & 0.9 & 18.063 & 3.35 $\pm$ 0.01 & 1.37 $\pm$ 0.03 & 44.75 $\pm$ 0.09 & 7.8 & $-$0.21 $\pm$ 0.1 \\
J0040$-$3713 & 0.4 & 15.969 & 3.22 $\pm$ 0.03 & 0.65 $\pm$ 0.04 & 42.86 $\pm$ 0.06 & 6.4 & $-$0.7 $\pm$ 0.09 \\
J0135$-$3539 & 3.0 & 17.888 & 3.03 $\pm$ 0.02 & 0.92 $\pm$ 0.03 & 44.08 $\pm$ 0.07 & 6.84 & 0.08 $\pm$ 0.09 \\
J0142$-$3525 & 0.5 & 17.733 & 3.04 $\pm$ 0.02 & 0.65 $\pm$ 0.03 & 43.88 $\pm$ 0.04 & 6.58 & 0.15 $\pm$ 0.06 \\
J0152$-$5238 & 1.0 & 17.557 & 3.18 $\pm$ 0.02 & 0.58 $\pm$ 0.04 & 43.54 $\pm$ 0.04 & 6.63 & $-$0.25 $\pm$ 0.06 \\
J0217$-$3004 & 0.5 & 17.227 & 3.2 $\pm$ 0.02 & 0.44 $\pm$ 0.05 & 42.97 $\pm$ 0.14 & 6.58 & $-$0.77 $\pm$ 0.15 \\
J0222$-$5332 & 2.6 & 17.983 & 3.24 $\pm$ 0.02 & 0.71 $\pm$ 0.04 & 43.74 $\pm$ 0.06 & 6.98 & $-$0.4 $\pm$ 0.07 \\
J0228$-$4057 & 1.6 & 17.623 & 3.24 $\pm$ 0.02 & 0.76 $\pm$ 0.01 & 45.83 $\pm$ 0.03 & 8.18 & 0.49 $\pm$ 0.07 \\
J0307$-$7250 & 0.1 & 15.359 & 3.35 $\pm$ 0.06 & 0.9 $\pm$ 0.04 & 42.73 $\pm$ 0.03 & 6.67 & $-$1.1 $\pm$ 0.12 \\
J0436$-$3010 & 4.0 & 18.026 & 3.14 $\pm$ 0.05 & 0.91 $\pm$ 0.2 & 43.54 $\pm$ 0.09 & 6.34 & 0.03 $\pm$ 0.15 \\
J0440$-$4110 & 1.2 & 17.19 & 2.94 $\pm$ 0.02 & 0.69 $\pm$ 0.02 & 43.25 $\pm$ 0.03 & 6.08 & 0.01 $\pm$ 0.06 \\
J0507$-$4652 & 10.0 & 18.934 & 2.95 $\pm$ 0.02 & 1.07 $\pm$ 0.06 & 43.55 $\pm$ 0.05 & 6.12 & 0.28 $\pm$ 0.07 \\
J0537$-$3817 & 1.2 & 18.811 & 3.2 $\pm$ 0.03 & 0.73 $\pm$ 0.05 & 44.42 $\pm$ 0.1 & 7.28 & $-$0.02 $\pm$ 0.13 \\
J0541$-$3738 & 1.1 & 18.121 & 3.06 $\pm$ 0.02 & 1.44 $\pm$ 0.04 & 44.37 $\pm$ 0.04 & 6.88 & 0.33 $\pm$ 0.06 \\
J0554$-$3453 & 0.6 & 17.903 & 3.08 $\pm$ 0.02 & 0.68 $\pm$ 0.04 & 43.75 $\pm$ 0.04 & 6.58 & 0.01 $\pm$ 0.06 \\
J0609$-$5606 & 0.5 & 16.731 & 2.84 $\pm$ 0.03 & 0.22 $\pm$ 0.03 & 42.99 $\pm$ 0.04 & 5.68 & 0.14 $\pm$ 0.08 \\
J0842$-$7042 & 1.0 & 17.38 & 3.34 $\pm$ 0.06 & 1.22 $\pm$ 0.1 & 43.85 $\pm$ 0.06 & 7.0 & $-$0.3 $\pm$ 0.13 \\
J1032$-$2707 & 1.1 & 16.715 & 3.09 $\pm$ 0.02 & 0.88 $\pm$ 0.03 & 43.7 $\pm$ 0.04 & 6.57 & $-$0.03 $\pm$ 0.07 \\
J1042$-$4000 & 19.2 & 18.845 & 3.22 $\pm$ 0.03 & 0.94 $\pm$ 0.04 & 44.83 $\pm$ 0.05 & 7.44 & 0.24 $\pm$ 0.09 \\
J1048$-$3902 & 0.5 & 16.757 & 3.26 $\pm$ 0.01 & 0.28 $\pm$ 0.02 & 43.42 $\pm$ 0.02 & 6.86 & $-$0.6 $\pm$ 0.04 \\
J1057$-$4039 & 621.1 & 18.006 & 3.21 $\pm$ 0.01 & 1.0 $\pm$ 0.02 & 45.27 $\pm$ 0.04 & 7.67 & 0.44 $\pm$ 0.06 \\
J1257$-$4435 & 2.1 & 16.731 & 2.87 $\pm$ 0.04 & 1.65 $\pm$ 0.06 & 43.92 $\pm$ 0.04 & 6.07 & 0.68 $\pm$ 0.1 \\
J1321$-$3104 & 0.9 & 17.573 & 3.33 $\pm$ 0.03 & 0.47 $\pm$ 0.06 & 43.03 $\pm$ 0.06 & 6.65 & $-$0.77 $\pm$ 0.1 \\
J1325$-$3824 & 1.7 & 16.387 & 2.8 $\pm$ 0.01 & 0.68 $\pm$ 0.03 & 44.43 $\pm$ 0.01 & 6.21 & 1.06 $\pm$ 0.04 \\
J1329$-$2925 & 6.3 & 19.388 & 3.25 $\pm$ 0.01 & 0.44 $\pm$ 0.01 & 44.5 $\pm$ 0.05 & 7.44 & $-$0.1 $\pm$ 0.06 \\
J1407$-$3120 & 0.9 & 17.052 & 2.95 $\pm$ 0.06 & 1.14 $\pm$ 0.08 & 43.25 $\pm$ 0.11 & 6.13 & $-$0.04 $\pm$ 0.16 \\
J1410$-$3644 & 0.6 & 16.659 & 3.15 $\pm$ 0.02 & 0.3 $\pm$ 0.02 & 43.93 $\pm$ 0.02 & 6.74 & 0.03 $\pm$ 0.05 \\
J1434$-$4254 & 7.9 & 17.92 & 3.04 $\pm$ 0.03 & 0.28 $\pm$ 0.05 & 43.52 $\pm$ 0.08 & 6.51 & $-$0.15 $\pm$ 0.1 \\
J1447$-$3802 & 0.4 & 16.948 & 3.27 $\pm$ 0.02 & 0.23 $\pm$ 0.02 & 43.42 $\pm$ 0.04 & 6.89 & $-$0.63 $\pm$ 0.06 \\
J1500$-$7248 & 112.1 & 17.945 & 3.18 $\pm$ 0.02 & 0.69 $\pm$ 0.03 & 44.09 $\pm$ 0.04 & 7.01 & $-$0.08 $\pm$ 0.07 \\
J1512$-$3333 & 0.8 & 15.152 & 3.22 $\pm$ 0.02 & 0.97 $\pm$ 0.04 & 42.68 $\pm$ 0.03 & 6.2 & $-$0.68 $\pm$ 0.06 \\
J1515$-$7820 & 6.5 & 17.134 & 3.2 $\pm$ 0.01 & 0.28 $\pm$ 0.01 & 44.82 $\pm$ 0.03 & 7.58 & 0.08 $\pm$ 0.05 \\
J1729$-$6854 & 0.5 & 16.525 & 3.34 $\pm$ 0.07 & 0.91 $\pm$ 0.05 & 43.47 $\pm$ 0.04 & 6.97 & $-$0.66 $\pm$ 0.15 \\
J1938$-$4326 & 0.9 & 16.995 & 3.25 $\pm$ 0.01 & 0.81 $\pm$ 0.01 & 43.86 $\pm$ 0.03 & 7.17 & $-$0.46 $\pm$ 0.04 \\
J2039$-$3018 & 5.0 & 17.573 & 3.05 $\pm$ 0.02 & 0.95 $\pm$ 0.05 & 43.38 $\pm$ 0.06 & 6.36 & $-$0.14 $\pm$ 0.08 \\
J2045$-$3010 & 0.9 & 18.105 & 3.21 $\pm$ 0.03 & 0.59 $\pm$ 0.05 & 43.82 $\pm$ 0.07 & 6.94 & $-$0.29 $\pm$ 0.1 \\
J2059$-$3147 & 0.9 & 16.42 & 2.85 $\pm$ 0.0 & 1.23 $\pm$ 0.03 & 43.66 $\pm$ 0.04 & 6.18 & 0.32 $\pm$ 0.05 \\
J2133$-$3558 & 0.6 & 18.235 & 3.25 $\pm$ 0.02 & 0.85 $\pm$ 0.04 & 44.47 $\pm$ 0.13 & 7.44 & $-$0.13 $\pm$ 0.14 \\
J2157$-$3123 & 1.0 & 17.902 & 3.06 $\pm$ 0.03 & 0.63 $\pm$ 0.04 & 43.27 $\pm$ 0.06 & 6.36 & $-$0.25 $\pm$ 0.09 \\
J2216$-$3947 & 2.2 & 18.351 & 3.24 $\pm$ 0.03 & 0.91 $\pm$ 0.03 & 44.31 $\pm$ 0.06 & 7.31 & $-$0.17 $\pm$ 0.09 \\
J2245$-$4652 & 3.8 & 16.991 & 3.36 $\pm$ 0.01 & 0.33 $\pm$ 0.01 & 45.2 $\pm$ 0.02 & 8.1 & $-$0.07 $\pm$ 0.05 \\
J2340$-$5941 & 5.7 & 18.686 & 3.19 $\pm$ 0.02 & 0.33 $\pm$ 0.03 & 44.31 $\pm$ 0.18 & 7.38 & $-$0.23 $\pm$ 0.18 \\
\hline
Median & 1.05 & 17.573 & 3.20 $\pm$ 0.02 & 0.72 $\pm$ 0.04 & 43.79 $\pm$ 0.04 & 6.79 & $-$0.09 $\pm$ 0.08 \\
\hline
\end{tabular}
\end{footnotesize}
\label{others}
\flushleft{\textbf{Notes.} Columns: (1) name, (2) radio loudness, (3) B-band magnitude, (4) FWHM of H$\beta$ line, (5) flux ratio of Fe~II / H$\beta$, (6) optical luminosity at 5100~$\text{\AA}$, (7) BH mass, (8) Eddington ratio. The last line shows the median values of the different parameters.}
\end{table}

\label{lastpage}
\end{document}